\documentclass[aps,prd,preprintnumbers,groupedaddress,nofootinbib,amssymb,eqsecnum,notitlepage]{revtex4-1}
\usepackage{graphicx}
\usepackage{color}
\usepackage{here}
\usepackage{amsmath,amsthm,amssymb}
\usepackage{bm}
\allowdisplaybreaks[1]

\usepackage{amsfonts}
\usepackage{dcolumn}
\usepackage[dvipdfmx]{hyperref}

\begin{document}

\newcommand{\newc}{\newcommand}
\newc{\tif}{\tilde{f}}
\newc{\tih}{\tilde{h}}
\newc{\tip}{\tilde{\phi}}
\newc{\tiA}{\tilde{A}}

\newc{\ben}{\begin{eqnarray}}
\newc{\een}{\end{eqnarray}}
\newc{\be}{\begin{equation}}
\newc{\ee}{\end{equation}}
\newc{\ba}{\begin{eqnarray}}
\newc{\ea}{\end{eqnarray}}
\newc{\bea}{\begin{eqnarray*}}
\newc{\eea}{\end{eqnarray*}}
\newc{\D}{\partial}
\newc{\ie}{{\it i.e.} }
\newc{\eg}{{\it e.g.} }
\newc{\etc}{{\it etc.} }
\newc{\etal}{{\it et al.}}
\newc{\nn}{\nonumber}
\newc{\ra}{\rightarrow}
\newc{\lra}{\leftrightarrow}
\newc{\lsim}{\buildrel{<}\over{\sim}}
\newc{\gsim}{\buildrel{>}\over{\sim}}
\newc{\aP}{\alpha_{\rm P}}
\newc{\dphi}{\delta\phi}
\newc{\da}{\delta A}
\newc{\tp}{\dot{\phi}}
\newc{\rhon}{\rho_{m,n}}
\newc{\rhonn}{\rho_{m,nn}}
\newc{\delj}{\delta j}
\newc{\delrho}{\delta \rho_m}
\newc{\pa}{\partial}
\newc{\aM}{\alpha_{\rm M}}
\newc{\aB}{\alpha_{\rm B}}
\newc{\E}{{\cal E}}

\title{Gauge-ready formulation of cosmological perturbations 
in scalar-vector-tensor theories}

\author{
Lavinia Heisenberg$^{1}$, 
Ryotaro Kase$^{2}$, and 
Shinji Tsujikawa$^{2}$}

\affiliation{
$^1$Institute for Theoretical Studies, ETH Zurich, 
Clausiusstrasse 47, 8092 Zurich, Switzerland\\
$^2$Department of Physics, Faculty of Science, 
Tokyo University of Science, 1-3, Kagurazaka,
Shinjuku-ku, Tokyo 162-8601, Japan}

\date{\today}

\begin{abstract}

In scalar-vector-tensor (SVT) theories with parity invariance, 
we perform a gauge-ready formulation of cosmological 
perturbations on the flat Friedmann-Lema\^{i}tre-Robertson-Walker 
(FLRW) background by taking into account a matter perfect fluid. 
We derive the second-order action of scalar perturbations 
and resulting linear perturbation equations of motion without 
fixing any gauge conditions. Depending on physical problems at hand, 
most convenient gauges can be chosen to study the development of 
inhomogeneities in the presence of scalar and vector fields coupled to 
gravity. This versatile framework, which encompasses Horndeski 
and generalized Proca theories as special cases, is applicable to 
a wide variety of cosmological phenomena including nonsingular 
cosmology, inflation, and dark energy. By deriving conditions for 
the absence of ghost and Laplacian instabilities in several 
different gauges, we show that, unlike Horndeski theories, 
it is possible to evade no-go arguments for the absence of 
stable nonsingular bouncing/genesis solutions in both generalized 
Proca and SVT theories. We also apply our framework to the 
case in which scalar and vector fields are responsible for 
dark energy and find that the separation of observables relevant to the 
evolution of matter perturbations into tensor, vector, and scalar sectors 
is transparent in the unitary gauge. Unlike the flat gauge chosen 
in the literature, this result is convenient to confront 
SVT theories with observations associated with the 
cosmic growth history.

\end{abstract}


\maketitle

\section{Introduction}
\label{introsec}

The cosmological perturbation theory is a fundamental framework for understanding the growth of cosmic
structures \cite{Bardeen,KS,MBran,BTW}.
The perturbations of metric and matter can be generally 
decomposed into scalar, vector, and tensor sectors arising from irreducible representations of the $SO(3)$ background field configuration. 
Among them, scalar perturbations are the main source for 
the development of inhomogeneities in the Universe.
For example, it is believed that the energy density of 
a scalar degree of freedom (DOF) drives inflation \cite{inf}, 
during which the field perturbation $\delta \phi$ 
is stretched over the Hubble radius \cite{oldper}. 
After inflation, the primordial curvature perturbation is converted 
to the radiation perturbation, which is observed as 
temperature anisotropies in the Cosmic Microwave 
Background (CMB) \cite{CMB}. 
The CMB temperature fluctuation works as a source for 
the growth of matter density perturbations due to 
the gravitational instability \cite{LSS}.

General Relativity (GR) with standard matter 
(baryons and radiation) is not sufficient to account for 
the observed evidence of inflation, dark energy, dark matter etc. 
It is possible to explain such phenomena by taking into account 
new DOFs like scalar or vector fields. 
As in the case of string dilaton \cite{Gas}, these new DOFs can 
have direct couplings to the gravity sector with two tensor polarizations.
For a single scalar field $\phi$ coupled to gravity, most general 
scalar-tensor theories with second-order equations of motion 
are known as Horndeski theories \cite{Horndeski,Horn1,Horn2,Horn3}. 
Indeed, the application of Horndeski theories to inflation and 
dark energy has been extensively performed in the 
literature \cite{Ginf,Burrage,DT11,GSami,DT10,deRham:2011by}.
Since different models in Horndeski theories predict different 
cosmic growth histories, one can distinguish them from the observations of CMB, redshift space distortions, 
weak lensing etc \cite{DKT,CCPS,Amendola12,Miguel,Bellini,Bellini2,Lom,Raveri,Peirone}.

For a massive vector field $A_{\mu}$ with broken $U(1)$ gauge symmetry, 
one can also construct self-interactions and nonminimal couplings 
to gravity similar to those appearing in
Horndeski theories \cite{Heisenberg,Tasinato,Allys,Jimenez}. 
The vector-tensor theories with second-order equations of motion 
are dubbed generalized Proca (GP) theories (see 
Refs.~\cite{HKT16,Kimura,Allys2,Jimenez17,Naka,KKNY} 
for further extensions).
If we apply GP theories to cosmology, the temporal 
vector component $A_0$ plays a role of the auxiliary field directly 
related to the Hubble expansion rate $H$ \cite{GPcosmo}. 
Then, there exists a de Sitter 
fixed point responsible for the late-time cosmic acceleration.

The important difference of GP theories from scalar-tensor theories is 
the presence of intrinsic vector modes in the former, 
which work as dynamical vector perturbations 
on the FLRW background. By choosing the flat gauge, the authors 
of Ref.~\cite{GPcosmo} obtained the 
second-order actions of scalar, vector, and tensor perturbations for 
the purpose of deriving stability conditions and observables relevant to 
the cosmic growth history. Existence of intrinsic vector modes affects 
the effective gravitational coupling with matter through a quantity 
$q_v$ associated with the no-ghost condition of vector 
perturbations \cite{Geff16}. 
Around local massive objects, nonlinear 
vector-field self-interactions can suppress the propagation of 
fifth forces through the operation of the Vainshtein 
mechanism \cite{GPsc1,GPsc2}.

In the presence of both scalar and vector fields coupled to gravity, it is 
possible to construct a unified version of Horndeski and GP theories 
with second-order equations of motion \cite{Heisenberg18}
(dubbed SVT theories). There are two versions of SVT theories, 
depending on whether the $U(1)$ gauge symmetry is respected or not.
The $U(1)$-invariant SVT theories have been already 
applied to the static and spherically symmetric configuration, 
in which case hairy black hole solutions endowed with scalar 
and vector hairs are present \cite{HT18,HKTodd,Pedro}.
The $U(1)$-broken SVT theories can be applied to the cosmological 
setup, in which the temporal vector component $A_0$ affects 
the background dynamics \cite{HKT18}. In this case, the longitudinal 
vector component works as a dynamical scalar perturbation.
In Ref.~\cite{Kasedark}, the second-order actions of tensor, vector, 
and scalar perturbations were derived in $U(1)$-broken SVT theories 
with parity invariance by choosing the flat gauge \cite{Kasedark}.
These results can be used for the studies of linear perturbations 
during inflation and late-time cosmic acceleration. 
See Ref.~\cite{Heisenberg:2018vsk} for a recent review on the 
systematic approach to generalizations of GR, where the novel progress 
in constructing consistent field theories of gravity based on 
additional scalar, vector and tensor fields together 
with their cosmological implications is reviewed.

In this paper, without fixing any gauge conditions from the beginning, 
we derive the second-order actions of scalar perturbations and resulting 
linear perturbation equations of motion in $U(1)$-broken SVT theories with 
parity invariance by taking into account a matter perfect fluid. 
The motivation of such analysis is that, depending on the problems 
at hand, the gauge should be appropriately chosen.

If we choose the flat gauge and apply GP/SVT theories to the bouncing 
cosmology, for example, the quantity $q_s$ relevant to no-ghost 
conditions of scalar perturbations (given by Eq.~(5.32) of 
Ref.~\cite{Kasedark}) vanishes at the bounce ($H=0$). 
However, this comes from the choice of the 
inappropriate gauge in which 
the curvature perturbation ${\cal R}$ vanishes at $H=0$.
If we choose other proper gauges like the Newtonian gauge, 
neither $q_s$ nor ${\cal R}$ vanishes at $H = 0$.
For Horndeski bouncing solutions that reduce to Einstein gravity both before and after 
the violation of null energy conditions, there is an instant so called the $\gamma$-crossing 
during the transition from Einstein to Horndeski regimes \cite{Ijjas}.
The quantity $\gamma$, which reduces to $H$ in Einstein gravity \cite{Xue}, appears 
in the denominators of the Hamiltonian and momentum constraints 
in the process of eliminating the nondynamical lapse and shift perturbations from 
the second-order scalar action. In the flat gauge there is a coordinate singularity at $\gamma=0$, 
but this does not correspond to a physical singularity in that it can be regulated 
by choosing a proper time slicing (choice of the Newtonian gauge in this case) \cite{Ijjas}.
These facts show that the flat gauge is not suitable for 
describing the evolution of scalar perturbations across the bounce.

If the flat gauge is chosen for the computation of observables 
relevant to the growth of matter perturbations in the dark 
energy cosmology, the coefficients of second-order scalar action 
do not explicitly contain terms associated with the stability of 
tensor perturbations. This reflects the fact that, unlike tensor 
perturbations, there are no scalar perturbations arising from 
spatial metric components in the flat gauge. 
In other gauges like the unitary gauge, 
we show that the second-order scalar action contains 
quantities related to stability conditions of tensor, 
vector, and scalar perturbations.
Then, unlike the flat gauge, the separation between tensor, vector, 
and scalar modes in the effective gravitational coupling $G_{\rm eff}$ 
of linear perturbations becomes transparent. 
This is convenient for testing dark energy models 
in SVT theories with observational data of the cosmic growth history. 

Our gauge-ready formulation of cosmological perturbations is versatile 
in that any convenient gauge can be chosen depending on 
the problem under consideration 
(see Sec.~12.1 in Ref.~\cite{Heisenberg:2018vsk} 
for further discussion on the gauge choice). 
We would like to stress that, provided the gauge is suitably chosen, 
physical results are equivalent to each other among different gauges.
For example, the effective gravitational coupling mentioned above can 
be expressed in several different ways by choosing different gauges, 
but they are actually identical to each other.
If the flat gauge is chosen from the beginning, 
expressing $G_{\rm eff}$ in terms of quantities associated with the stability 
conditions of tensor perturbations is a nontrivial and complicated procedure.
This is attributed to the mixture 
of those quantities among coefficients of the second-order action of 
scalar perturbations.
This problem can be avoided in our gauge-ready formalism in which 
the gauge choice can be performed at the level of  
scalar perturbation equations of motion. 
Apart from a subclass of Horndeski 
theories \cite{Hwang}, this gauge-ready formulation was not 
performed yet even for full Horndeski theories.
Our results are sufficiently general to accommodate both 
Horndeski and GP theories as specific cases. 

Our paper is organized as follows.
In Sec.~\ref{modelsec}, we revisit the background 
equations of motion in SVT theories as well as the second-order actions 
of tensor and vector perturbations.
In Sec.~\ref{scasec}, we derive the second-order action of scalar 
perturbations and resulting perturbation equations of motion without 
fixing any gauges. We also discuss the issues of gauge transformations, 
gauge-invariant variables, and gauge choices.
In Sec.~\ref{stasec}, we obtain conditions for the absence of 
ghost and Laplacian instabilities of scalar perturbations 
in the small-scale limit 
by choosing several different gauges.
In Sec.~\ref{appsec}, our general results are applied 
to the discussion for the realization of stable nonsingular 
bouncing/genesis cosmologies. 
In Sec.~\ref{conmodelsec}, we compute observables relevant to 
the evolution of Newtonian and weak lensing gravitational 
potentials by choosing the unitary gauge in scalar perturbation 
equations of motion.
Sec.~\ref{concludesec} is devoted to conclusions.

\section{SVT theories on the cosmological background}
\label{modelsec}

In SVT theories with broken $U(1)$ gauge symmetry \cite{Heisenberg18}, 
there exist a scalar field $\phi$ and a vector field $A_{\mu}$ coupled to gravity.  
For the vector field, we define the antisymmetric field strength tensor 
$F_{\mu \nu}$, its dual $\tilde{F}_{\mu \nu}$, and the symmetric 
tensor $S_{\mu\nu}$, as
\be
F_{\mu\nu}=\nabla_\mu A_\nu-\nabla_\nu A_\mu\,,
\qquad 
\tilde{F}^{\mu\nu}=\frac{1}{2}
\mathcal{E}^{\mu\nu\alpha\beta}F_{\alpha\beta}\,,
\qquad 
S_{\mu \nu}=\nabla_{\mu}A_{\nu}
+\nabla_{\nu}A_{\mu}\,,
\ee
where $\nabla_{\mu}$ represents the covariant derivative operator 
and $\mathcal{E}^{\mu\nu\alpha\beta}$ is the 
antisymmetric Levi-Civita tensor.
The SVT theories contain the following Lorentz-invariant 
combinations:
\be
X_1=-\frac{1}{2} \nabla_{\mu} \phi \nabla^{\mu} \phi\,,
\qquad
X_2=-\frac{1}{2} A^{\mu} \nabla_{\mu} \phi \,,
\qquad 
X_3=-\frac{1}{2} A_{\mu} A^{\mu}\,,
\label{X123}
\ee
and
\be
F=-\frac{1}{4}F_{\mu\nu}F^{\mu\nu}\,,\qquad
Y_1=\nabla_\mu \phi \nabla_\nu \phi\,F^{\mu\alpha}F^\nu{}_\alpha\,,
\qquad 
Y_2=\nabla_\mu\phi\, A_\nu F^{\mu\alpha}F^\nu{}_\alpha\,, 
\qquad
Y_3=A_\mu A_\nu F^{\mu\alpha}F^\nu{}_\alpha\,.
\label{FY123}
\ee
The quantities $X_1$ and $X_3$ correspond to the kinetic term 
of $\phi$ and the mass term of $A_{\mu}$, respectively, while 
$X_2$ characterizes their mixings. The quantities 
$F, Y_1, Y_2, Y_3$ arise from intrinsic vector modes.

The Ricci scalar $R$ and Einstein tensor $G_{\mu \nu}$ 
are generally coupled to scalar and vector fields. 
To keep the equations of motion up to second order, 
we need to take into account additional derivative 
interactions of those fields.
In SVT theories, there are also nonminimal 
couplings with the double dual Riemann tensor defined by 
\be
L^{\mu\nu\alpha\beta}=\frac{1}{4}
\mathcal{E}^{\mu\nu\rho\sigma}
\mathcal{E}^{\alpha\beta\gamma\delta} 
R_{\rho\sigma\gamma\delta}\,,
\ee
where $R_{\rho\sigma\gamma\delta}$ is the Riemann tensor. 

\subsection{Action of SVT theories with 
broken $U(1)$ gauge invariance}

The full action of parity-invariant SVT theories with broken 
$U(1)$ gauge invariance is given by  \cite{Heisenberg18}
\be
\mathcal{S}=
\int d^4x \sqrt{-g}\, \left( 
\sum_{n=3}^5 
\mathcal{L}_{{\rm ST}}^{(n)}
+\sum_{n=2}^6
\mathcal{L}_{{\rm SVT}}^{(n)}
\right)+{\cal S}_m\,,
\label{action}
\ee
where $g$ is the determinant of metric tensor $g_{\mu \nu}$.
The Lagrangians $\mathcal{L}_{{\rm ST}}^{(n)}$ and 
$\mathcal{L}_{{\rm SVT}}^{(n)}$ are those arising 
in scalar-tensor (Horndeski) theories and 
SVT theories, respectively, 
whose explicit forms are 
\ba
{\cal L}_{\rm ST}^{(3)} &=& 
G_3(\phi,X_1) \square \phi\,,\\
{\cal L}_{\rm ST}^{(4)} &=& 
G_4(\phi,X_1)R+G_{4,X_1}(\phi,X_1) 
\left[ (\square \phi)^2 -(\nabla_{\mu}\nabla_{\nu}\phi) 
(\nabla^{\mu}\nabla^{\nu}\phi) \right]\,,\\
{\cal L}_{\rm ST}^{(5)} 
&=&
G_5(\phi,X_1) G_{\mu \nu} (\nabla^{\mu} \nabla^{\nu} \phi) \nonumber \\
& &
-\frac{1}{6}G_{5,X_1}(\phi,X_1) \left[ (\square \phi)^3 
-3  (\square \phi) (\nabla_{\mu}\nabla_{\nu}\phi) 
(\nabla^{\mu}\nabla^{\nu}\phi)
+2(\nabla^{\mu}\nabla_{\alpha}\phi ) 
(\nabla^{\alpha}\nabla_{\beta}\phi ) 
(\nabla^{\beta}\nabla_{\mu}\phi )\right]\,,
\label{STaction}
\ea
and
\ba
\mathcal{L}_{{\rm SVT}}^{(2)} &=&
f_2(\phi,X_1,X_2,X_3,F,Y_1,Y_2,Y_3)\,, \\
\mathcal{L}_{{\rm SVT}}^{(3)}  &=&
f_{3}(\phi,X_3)g^{\mu\nu}S_{\mu\nu}
+\tilde{f}_{3}(\phi,X_3)A^{\mu}A^{\nu} S_{\mu\nu}\,,\\
\mathcal{L}_{{\rm SVT}}^{(4)} & = & 
f_{4}(\phi,X_3)R+f_{4,X_3}(\phi,X_3) \left[ 
(\nabla_\mu A^\mu)^2-\nabla_\mu A_\nu \nabla^\nu A^\mu \right]\,, \\
\mathcal{L}_{{\rm SVT}}^{(5)} & = & 
f_5(\phi,X_3)G^{\mu\nu} \nabla_{\mu}A_{\nu}
-\frac{1}{6}f_{5,X_3}(\phi,X_3) 
\left[ (\nabla_{\mu} A^{\mu})^3
-3\nabla_{\mu} A^{\mu}
\nabla_{\rho}A_{\sigma} \nabla^{\sigma}A^{\rho}
+2\nabla_{\rho}A_{\sigma} \nabla^{\gamma}
A^{\rho} \nabla^{\sigma}A_{\gamma}\right] \nonumber \\
&+&\mathcal{M}_5^{\mu\nu}\nabla_\mu \nabla_\nu\phi
+\mathcal{N}_5^{\mu\nu}S_{\mu\nu}\,, \\
\mathcal{L}_{{\rm SVT}}^{(6)} & = &
f_6(\phi,X_1)L^{\mu\nu\alpha\beta}F_{\mu\nu}
F_{\alpha\beta}
+2f_{6,X_1} (\phi, X_1) 
\tilde{F}^{\mu\nu}\tilde{F}^{\alpha\beta}\nabla_\mu\nabla_\alpha \phi\nabla_\nu
\nabla_\beta\phi \nonumber \\
& &
+\tilde{f}_6(\phi,X_3)L^{\mu\nu\alpha\beta}F_{\mu\nu}F_{\alpha\beta}+
\frac12\tilde{f}_{6,X_3} (\phi, X_3) 
\tilde{F}^{\mu\nu}\tilde{F}^{\alpha\beta}
S_{\mu \alpha}S_{\nu\beta}\,,
\label{LagSVT}
\ea
with the notations $\square \phi=g^{\mu \nu} 
\nabla_{\mu} \nabla_{\nu} \phi$ and 
$G_{i,X_1}=\partial G_i/\partial X_1$, 
$f_{i,X_3}=\partial f_i/\partial X_3$ etc. 
The functions $G_3, G_4, G_5$ depend on $\phi$ and 
its kinetic energy $X_1$. 
The quadratic Horndeski Lagrangian $G_2(\phi, X_1)$ 
is accommodated in the SVT Lagrangian 
$f_2$, which is a function of  
$\phi, X_i, F, Y_i$ (where $i=1,2,3$).
The functions $f_3, \tilde{f}_3, f_4, f_5, \tilde{f}_6$ 
depend on $\phi$ and $X_3$, while $f_6$ 
is a function of $\phi$ and $X_1$. 

The 2-rank tensors $\mathcal{M}^{\mu\nu}_5$ and 
$\mathcal{N}^{\mu\nu}_5$ in ${\cal L}_{\rm SVT}^{(5)}$ 
are defined, respectively,  by 
\be
\mathcal{M}^{\mu\nu}_5
=\mathcal{G}_{\rho\sigma}^{h_{5}} 
\tilde{F}^{\mu\rho}\tilde{F}^{\nu\sigma}\,,\qquad 
\mathcal{N}^{\mu\nu}_5
=\mathcal{G}_{\rho\sigma}^{\tilde{h}_{5}} 
\tilde{F}^{\mu\rho}\tilde{F}^{\nu\sigma}\,,
\ee
where
\ba
\mathcal{G}_{\rho\sigma}^{h_5} 
&=& 
h_{51}(\phi,X_i)g_{\rho\sigma}+h_{52}(\phi,X_i)
\nabla_\rho \phi \nabla_\sigma \phi
+h_{53}(\phi,X_i)A_\rho A_\sigma
+h_{54}(\phi,X_i)A_\rho \nabla_\sigma \phi\,,\\
\mathcal{G}_{\rho\sigma}^{\tilde{h}_5} 
&=& 
\tilde{h}_{51}(\phi,X_i)g_{\rho\sigma}
+\tilde{h}_{52}(\phi,X_i)
\nabla_\rho \phi \nabla_\sigma \phi
+\tilde{h}_{53}(\phi,X_i)A_\rho A_\sigma
+\tilde{h}_{54}(\phi,X_i)A_\rho \nabla_\sigma \phi\,,
\label{effmet}
\ea
with the functions $h_{5j}$ and $\tilde{h}_{5j}$ 
($j=1,2,3,4$) depending on $\phi$ and $X_i$.
The Lagrangians $\mathcal{M}_5^{\mu\nu}\nabla_\mu 
\nabla_\nu\phi$, $\mathcal{N}_5^{\mu\nu}S_{\mu\nu}$, 
and ${\cal L}_{\rm SVT}^{(6)}$
correspond to intrinsic vector modes. 
The last two terms in Eq.~(\ref{LagSVT}) 
appear in GP theories with the $X_3$ dependence alone in $\tilde{f}_6$. 
The first two terms in Eq.~(\ref{LagSVT}) and the 
$\phi$ dependence in $\tilde{f}_6$ arise in the context 
of SVT theories.
As pointed out in Ref.~\cite{Heisenberg18}, the full dependence 
of tensors $\mathcal{M}_5^{\mu \nu}$ and 
$\mathcal{N}_5^{\mu \nu}$ on all the functions 
$h_{5j}$ and $\tilde{h}_{5j}$ in the effective metric would introduce dynamics for the temporal component of the vector field 
on a general background and hence an additional restriction 
is needed. To guarantee the absence of ghosts on arbitrary backgrounds, 
the dependence of $\mathcal{M}_5^{\mu \nu}$ has to be restricted 
to $X_1$ only and similarly the dependence of 
$\mathcal{N}_5^{\mu \nu}$ to $X_3$, but for the purpose of 
cosmological applications, we keep the analysis general here. 

In Eq.~(\ref{action}), we have taken into account the matter 
action ${\cal S}_m$ to include additional DOFs like radiation,  
dark matter, and baryons. 
For this matter sector, we consider a perfect fluid minimally 
coupled to gravity.

\subsection{Background equations of motion}

We consider the flat FLRW background given by 
the line element
\be
ds^2=-dt^2+a^2(t) \delta_{ij}dx^i dx^j\,,
\label{FLRW}
\ee
where $a(t)$ is the time-dependent scale factor.  
The Hubble expansion rate is defined by 
$H(t)=\dot{a}(t)/a(t)$, where a dot represents 
a derivative with respect to $t$. 
The scalar and vector fields compatible with the background 
(\ref{FLRW}) are of the forms $\phi=\phi(t)$ and  
$A_{\mu}(t)=\left( A_0(t), 0, 0, 0 \right)$, 
where the temporal component $A_0(t)$ corresponds 
to a time-dependent auxiliary field.  
The matter sector is described by a perfect fluid 
with energy density $\rho_m$ and pressure $P_m$.

The background equations of motion on the flat FLRW 
spacetime (\ref{FLRW}) were already derived 
in Ref.~\cite{Kasedark} (see also Ref.~\cite{HKT18}).
By using coefficients of the second-order action of 
scalar perturbations, they can be expressed 
in compact forms, as 
\ba
& &
6\left( f_4+G_4 \right)H^2 +f_2 -\dot{\phi}^2 f_{2,X_1}-\frac{1}{2} 
\dot{\phi}A_0 f_{2,X_2}+\dot{\phi}^2 \left( 3H \dot{\phi}G_{3,X_1} 
-G_{3,\phi} \right)+6H \left( \dot{\phi}f_{4,\phi}
-HA_0^2 f_{4,X_3} \right) \nonumber \\
& &
+6 H \dot{\phi} \left( G_{4,\phi}+\dot{\phi}^2 G_{4,X_1 \phi}
-2H \dot{\phi}G_{4,X_1}-H \dot{\phi}^3 G_{4,X_1 X_1} \right)
+2A_0H^2 \left( 3\dot{\phi}f_{5,\phi}-H A_0^2 f_{5,X_3} \right)  \nonumber \\
& &
+H^2 \dot{\phi}^2 \left( 9 G_{5,\phi}+3\dot{\phi}^2 G_{5,X_1 \phi} 
-5H \dot{\phi}G_{5,X_1}-H \dot{\phi}^3 G_{5,X_1 X_1} \right)=\rho_m\,,
\label{back1}\\
& &
2q_t \dot{H}-D_6 \ddot{\phi}+\frac{w_2}{A_0} \dot{A}_0+D_7 \dot{\phi}
=-\rho_m-P_m\,,
 \label{back2}\\
& & 
3D_6 \dot{H}+2D_1 \ddot{\phi}-D_8 \dot{A}_0
+3D_7 H-D_9 A_0-D_5=0\,,
 \label{back3}\\
& &
2 \left( f_{2,X_3}+6H^2f_{4,X_3}-6H \dot{\phi} f_{4,X_3 \phi} 
\right)A_0-2 \left( 6H f_{3,X_3}+6H \tilde{f}_3
+2\dot{\phi}\tilde{f}_{3,\phi}-3H^3 f_{5,X_3}
+3H^2 \dot{\phi}f_{5,X_3 \phi} \right)A_0^2 \nonumber \\
\hspace{-0.3cm}
& &
+12H^2f_{4,X_3X_3} A_0^3 +2H^3 f_{5,X_3X_3} A_0^4
+\left( f_{2,X_2}+4f_{3,\phi}-6H^2 f_{5,\phi} \right)
\dot{\phi}=0\,,
 \label{back4} \\
& &
\dot{\rho}_m+3H \left( \rho_m+P_m \right)=0\,,
\label{coneq}
\ea
where $D_1,D_5,D_6,D_7,D_8,D_9$ and $w_2$ 
are given in Appendix \ref{coeff}. 
The quantity $q_t$ in Eq.~(\ref{back2}) is defined by 
\be
q_t=2f_4+2G_4-2A_0^2 f_{4,X_3}-2\dot{\phi}^2 G_{4,X_1}
+A_0 \dot{\phi} f_{5,\phi}-HA_0^3 f_{5,X_3}
+\dot{\phi}^2 G_{5,\phi}-H \dot{\phi}^3 G_{5,X_1}\,,
\label{qt}
\ee
whose positivity is required for the absence of ghosts 
in the tensor sector (see Sec.~\ref{tvsec}). 
We note that Eqs.~(\ref{back1})-(\ref{back2}) follow from 
Hamiltonian and momentum constraints, whereas 
Eqs.~(\ref{back3}), (\ref{back4}), and (\ref{coneq}) 
correspond to the equations of motion for $\phi$, $A_0$, 
and the perfect fluid, respectively.
Differentiating Eq.~(\ref{back4}) with respect to $t$, 
it follows that 
\be
\frac{2w_5}{A_0^2} \dot{A}_0-D_8 \ddot{\phi}
-\frac{3w_2}{A_0} \dot{H}
-D_9 \dot{\phi}=0\,,
\label{back5}
\ee
where $w_5$ is given in Appendix \ref{coeff}. 
Then, we can solve Eqs.~(\ref{back2}), (\ref{back3}), 
and (\ref{back5}) 
for $\dot{A}_0$, $\ddot{\phi}$, and $\dot{H}$ 
under the condition 
\be
{\cal D} \equiv 2 \left( 4D_1q_t w_5+3D_1w_2^2
+3D_6^2w_5-A_0^2 D_8^2 q_t-3A_0D_6D_8 w_2
\right) \neq 0\,.
\label{calD}
\ee
The determinant ${\cal D}$ cannot change its sign to 
avoid divergences of the quantities $\dot{A}_0$, 
$\ddot{\phi}$, $\dot{H}$.
Indeed, ${\cal D}$ is proportional to a quantity 
$q_s$ associated with the no-ghost condition 
of scalar perturbations \cite{Kasedark}, 
so that the positivity of $q_s$ corresponds 
to ${\cal D}>0$.

\subsection{Stability conditions of tensor and 
vector perturbations}
\label{tvsec}

The conditions for the absence of ghost and Laplacian 
instabilities of tensor perturbations $h_{ij}$ 
were derived in Ref.~\cite{Kasedark}.
The perturbed line element in the tensor sector 
is given by 
\be
ds_t^2=-dt^2+a^2(t) \left( \delta_{ij}+h_{ij} 
\right) dx^i dx^j\,,
\ee
where the nonvanishing components of $h_{ij}$ 
can be chosen as 
$h_{11}=h_1(t,z)$, 
$h_{22}=-h_1(t,z)$, and 
$h_{12}=h_{21}=h_2(t,z)$
to satisfy the transverse and traceless conditions
$\partial^j h_{ij}=0$ and ${h_i}^i=0$.
Expanding the action (\ref{action}) in terms of $h_{ij}$ up to 
quadratic order, the resulting second-order action of
tensor perturbations yields
\be
{\cal S}_t^{(2)}=\int dt d^3x \sum_{i=1}^{2}
\frac{a^3}{4}q_t \left[ \dot{h}_i^2-\frac{c_t^2}{a^2} 
(\partial h_i)^2 \right]\,,
\label{St2}
\ee
where $q_t$ is given by Eq.~(\ref{qt}), and
\be
c_t^2=\frac{1}{q_t} 
\left( 2f_4+2G_4-A_0\dot{\phi}f_{5,\phi}
-\dot{A}_0 A_0^2 f_{5,X_3}-\dot{\phi}^2G_{5,\phi} 
-\dot{\phi}^2 \ddot{\phi} G_{5,X_1} \right)\,.
\label{ct}
\ee
Since we are considering the theories with a massless graviton, 
the term proportional to $h_i^2$ in the second-order action 
vanishes after the integration by parts. 
We require the two conditions $q_t>0$ and $c_t^2>0$ to 
avoid ghost and Laplacian instabilities.

For vector perturbations, the perturbed line element 
in the flat gauge is given by 
\be
ds_v^2=-dt^2+2V_i dt dx^i+a^2(t) \delta_{ij}
dx^i dx^j\,,
\ee
where $V_i$ satisfies the transverse condition 
$\partial^i V_i=0$. The spatial components of 
$A_{\mu}$ can be expressed as $A_i=Z_i+\partial_i \psi$, 
where $Z_i$ is the vector perturbation obeying 
$\partial^i Z_i=0$ and $\psi$ is the longitudinal scalar 
perturbation discussed later in Sec.~\ref{scasec}. 
For the components of $Z_i$, we choose 
$Z_i=(Z_1(t,z), Z_2(t,z), 0)$ without loss of generality. 
The matter perfect fluid can be described by 
a Schutz-Sorkin action \cite{Sorkin,DGS}, see Eq.~(2.16) 
of Ref.~\cite{Kasedark}. 
However, it gives rise to only nondynamical perturbations 
like the velocity perturbation $v_i$.
After integrating out all the nondynamical perturbations
and taking the small-scale limit, we are left with two 
dynamical DOFs $Z_1$ and $Z_2$ with the quadratic
action \cite{Kasedark}
\be
{\cal S}_v^{(2)}=\int dt d^3 x \sum_{i=1}^2 
\frac{a}{2}q_v \left[ \dot{Z}_i^2-\frac{c_v^2}{a^2} 
\left( \partial Z_i \right)^2-\frac{\alpha_2}{q_v}Z_i^2 
\right]\,,
\label{Sv2}
\ee
where 
\ba
\hspace{-0.5cm}
q_v &=& f_{2,F}+2\dot{\phi}^2 f_{2,Y_1}
+2\dot{\phi}A_0 f_{2,Y_2}+2A_0^2f_{2,Y_3}
-4H \left( \dot{\phi} h_{51}+2A_0  \tilde{h}_{51} \right)
+8H^2 \left( f_6+\tilde{f}_6+\dot{\phi}^2 f_{6,X_1}
+A_0^2 \tilde{f}_{6,X_3} \right)\,,\label{qv}\\
\hspace{-0.5cm}
c_v^2 &=&
\frac{2\alpha_1 q_t+\alpha_3^2}{2q_t q_v}\,,
\label{cv}
\ea
with
\ba
\alpha_1 &=& f_{2,F} - 4\dot{A}_0\tilde{h}_{51} 
+8\left( H^2+\dot{H} \right) \left( f_6+\tilde{f}_6 \right)
-2\ddot{\phi}\,h_{51}
+H \biggl[ 2\dot{\phi} \left( \dot{\phi}^2 h_{52}
-h_{51}+4\ddot{\phi}f_{6,X_1} \right)  \nonumber \\
& &
-A_0 \left\{
4\tilde{h}_{51}-2\dot{\phi}^2
\left( h_{54}+2\tilde{h}_{52} \right) 
-8\dot{A}_0 \tilde{f}_{6,X_3}\right\}
+2\dot{\phi}A_0^2(h_{53}+2\tilde{h}_{54})
+4A_0^3 \tilde{h}_{53}
\biggr]\,,\\
\alpha_2 &=& -w_7 \nonumber \\
&=&
f_{2,X_3}+4\dot{H}f_{4,X_3}
-2 \left( \dot{A}_0+3HA_0 \right) 
\left( f_{3,X_3}+\tilde{f}_3 \right)
-2\dot{\phi}A_0 \tilde{f}_{3,\phi}
+2H ( 3H f_{4,X_3}+
3HA_0^2 f_{4,X_3 X_3}+2A_0 \dot{A}_0
f_{4,X_3X_3} \nonumber \\
& &-\dot{\phi}f_{4,X_3 \phi})
+H \left( H \dot{A}_0+2 \dot{H} A_0+3H^2 A_0
\right) f_{5,X_3}+H^2 A_0 \left( 
H A_0^2 f_{5,X_3 X_3}+A_0 \dot{A}_0 
f_{5,X_3 X_3}-2\dot{\phi}f_{5,X_3 \phi}\right)\,,\\
\alpha_3 &=& 
-2A_0 f_{4,X_3}-HA_0^2 f_{5,X_3}+\dot{\phi}f_{5,\phi}\,.
\label{al3}
\ea
The quantity $w_7$, which has the opposite sign to $\alpha_2$, 
appears in the second-order action of scalar perturbations, 
see Appendix \ref{coeff}.
The term $\alpha_2$ is associated  
with the mass squared of vector perturbations.
Provided that $\alpha_2>0$, there is no tachyonic instability of 
vector perturbations. Even for $\alpha_2<0$, as long as 
the mass $\sqrt{-\alpha_2}$ is as light as today's Hubble constant $H_0$,
the tachyonic instability does not arise for perturbations 
inside the Hubble radius. 
In the small-scale limit, there are neither ghost nor Laplacian instabilities 
for $q_v>0$ and $c_v^2>0$, whose conditions are independent of 
the choice of gauges.

\section{Gauge-ready formulation of scalar perturbations}
\label{scasec}

In this section, we derive the second-order action of scalar 
perturbations without fixing gauge conditions. 
The resulting linear perturbation equations of motion are 
written in the gauge-ready form, so that one can choose 
convenient gauges depending on the problems at hand. 
Let us consider the perturbed line element containing 
four scalar metric perturbations $\alpha$, $\chi$ $\zeta$, 
and $E$ \cite{Bardeen}: 
\be
ds_s^2=-(1+2\alpha) dt^2+2 \partial_i \chi dt dx^i 
+a^2(t) \left[ (1+2\zeta) \delta_{ij}
+2\partial_{i}\partial_{j}E \right] dx^idx^j\,,
\label{line}
\ee
where 
$\partial_i \chi \equiv \partial \chi/\partial x^i$ and 
$\partial_i \partial_j E \equiv \partial^2 E/\partial 
x^i \partial x^j$. 
The scalar and vector fields are expressed in the forms
\ba
\phi&=&\bar{\phi}(t)+\delta \phi\,,\\
A^{0}&=&-\bar{A}_0(t)+\delta A\,,\qquad 
A_i=\partial_i \psi\,,
\ea
where $\bar{\phi}(t)$, $\bar{A}_0(t)$ are background quantities, 
$A_i$ is the spatial component of $A_{\mu}$,  
and $\delta \phi$, $\delta A$, $\psi$ are scalar perturbations. 
In the following, we omit the over-bar from 
background quantities. 

\subsection{Second-order matter action}

To describe scalar perturbations in the matter sector, we consider 
the matter perfect fluid described by the Schutz-Sorkin 
action \cite{Sorkin,DGS}:
\be
{\cal S}_m=-\int d^{4}x \left[ \sqrt{-g}\,\rho_m(n)
+J^{\mu} \partial_{\mu}\ell \right]\,.
\label{Spf}
\ee
The quantity $J^{\mu}$ is related to the number density 
$n$, as 
\be
n=\sqrt{\frac{J^{\mu}J^{\nu}g_{\mu \nu}}{g}}\,. 
\label{ndef}
\ee
The temporal and spatial components of $J^{\mu}$ can be 
decomposed into background and perturbed parts, as 
\ba
J^{0} = \mathcal{N}_{0}+\delta J\,,\qquad
J^{k} =\frac{1}{a^2(t)}\,\delta^{ki}\partial_{i}\delta j\,,
\label{elldef}
\ea
where $\mathcal{N}_{0}$ is a constant, and 
$\delta J, \delta j$ are scalar perturbations. 
The background number density $n_0$ is given by 
$n_0=\mathcal{N}_{0}/a^3$.
The scalar quantity $\ell$ has a relation to the 
velocity potential $v$, as   
\be
\ell=-\int^{t} \rho_{m,n} 
(\tilde{t})d\tilde{t} 
-\rho_{m,n}v\,, 
\label{ells}
\ee
where $\rho_{m,n}\equiv\partial\rho_{m}/\partial n$. 
We introduce the matter density perturbation $\delta \rho_m$ 
in the form 
\be
\delta \rho_m=\frac{\rho_{m,n}}{a^3} \left[\delta J
-\mathcal{N}_{0}(3\zeta+\partial^2E)\right]\,,
\ee
where we use the notation 
$\partial^2E \equiv (\partial_i E)(\partial_i E)$ 
with the same latin subscripts summed over. 
The perturbation of fluid number density $n$, up to second order, is given by 
\be
\delta n= \frac{\delta \rho_m}{\rho_{m,n}}
-\frac{({\cal N}_0 \partial \chi+\partial \delta j)^2}{2{\cal N}_0a^5}
-\frac{(3\zeta+\partial^2E)\delta \rho_m}{\rho_{m,n}}
-\frac{{\cal N}_0(\zeta+\partial^2E)(3\zeta-\partial^2E)}{2a^3}\,.
\ee
At linear order, this reduces to 
$\delta n=\delta \rho_m/\rho_{m,n}$.

Expanding the Schutz-Sorkin action (\ref{Spf}) up to quadratic 
order in scalar perturbations, it follows that 
\ba
({\cal S}_m^{(2)})_s
&=&\int dt d^3 x\, a^3
\Bigg[
\frac{\rhon}{2a^8n_0}(\pa \delj)^2+\frac{\rhon}{a^5}(\pa\chi+\pa v) (\pa\delj)
+\left(\dot{v}-3Hc_m^2v-\alpha\right)\delrho
-\frac{c_m^2}{2n_0\rhon}\delrho^2+\frac{\rho_m}{2}\alpha^2
\notag\\
&&
+\frac{n_0 \rhon-\rho_m}{2}\left\{\frac{(\pa\chi)^2}{a^2}
+(\zeta+\pa^2E)(3\zeta-\pa^2E)\right\}
+(3\zeta+\pa^2E)\left\{ n_0\rhon(\dot{v}-3Hc_m^2 v)-\rho_m\alpha\right\}
\Bigg]\,,
\label{SMS}
\ea
where $c_m^2$ is the matter sound speed squared defined by 
\be
c_m^2=\frac{P_{m,n}}{\rho_{m,n}}
=\frac{n_0 \rho_{m,nn}}{\rho_{m,n}}\,.
\ee
Varying Eq.~(\ref{SMS}) with respect to $\delj$, 
we obtain 
\be
\partial \delta j=-a^3 n_0 \left( \partial v+\partial \chi 
\right)\,.
\ee
Substituting this relation into Eq.~(\ref{SMS}), 
the second-order matter action reduces to 
\ba
\hspace{-0.3cm}
({\cal S}_m^{(2)})_s
&=&\int dt d^3x\,a^3 
\Bigg[ \left( \dot{v}-3H c_m^2v-\alpha 
\right) \delta \rho_m-\frac{c_m^2}{2n_0 \rho_{m,n}}\delrho^2
-\frac{n_0 \rho_{m,n}}{2a^2} 
\left\{ (\partial v)^2+2\partial v \partial \chi \right\}
-\frac{\rho_m}{2a^2} (\partial \chi)^2
+\frac{\rho_m}{2}\alpha^2
\notag\\
\hspace{-0.3cm}
&&
+\frac{P_m}{2}(\zeta+\pa^2E)(3\zeta-\pa^2E)
+(3\zeta+\pa^2E)\left\{n_0 \rhon(\dot{v}-3Hc_m^2 v)
-\rho_m\alpha\right\} \Bigg]\,,
\label{SMS2}
\ea
where we used the property that the background pressure 
is given by 
\be
P_m=n_0 \rho_{m,n}-\rho_m\,.
\ee
The second-order matter action (\ref{SMS2}) is written 
in a gauge-ready form.

\subsection{Full second-order action and 
perturbation equations of motion
in gauge-ready form}

Now, we expand the total action (\ref{action}) up to quadratic order 
in scalar perturbations. On using Eq.~(\ref{back1}), the term 
$\rho_m\alpha^2/2$ in Eq.~(\ref{SMS2}) is cancelled
by a part of contributions proportional to $\alpha^2$ arising from 
$\mathcal{L}_{{\rm ST}}^{(n)}+\mathcal{L}_{{\rm SVT}}^{(n)}$. 
After integrations by parts, the full second-order action 
is expressed in the form 
\be
{\cal S}_s^{(2)}=\int dt d^3x 
\left({\cal L}^{\rm flat}_{1}+{\cal L}^{\rm flat}_{2}+{\cal L}^{\rm flat}_{3}
+{\cal L}_{\zeta}+{\cal L}_{E}\right)\,, 
\label{Ss}
\ee
where 
\ba
\hspace{-0.7cm}
{\cal L}^{\rm flat}_{1}
&=& a^3\left[
D_1\dot{\dphi}^2+D_2\frac{(\partial\dphi)^2}{a^2}+D_3\dphi^2
+\left(D_4\dot{\dphi}+D_5\dphi+D_6\frac{\partial^2\dphi}{a^2}\right) \alpha
-\left(D_6\dot{\dphi}-D_7\dphi\right)\frac{\partial^2\chi}{a^2}
\right.\notag\\
\hspace{-0.7cm}
&&
\left.~~~
+\left(D_8\dot{\dphi}+D_9\dphi\right)\da+D_{10}\,\dphi\,\frac{\partial^2\psi}{a^2}
\right]\,, \label{L1}\\
\hspace{-0.7cm}
{\cal L}^{\rm flat}_{2}
&=& a^3\left[
\left(w_1\alpha-w_2\frac{\da}{A_0}\right)\frac{\partial^2\chi}{a^2}
-w_3\frac{(\partial\alpha)^2}{a^2}+w_4\alpha^2
-\left(w_3\frac{\partial^2\da}{a^2A_0}-w_8\frac{\da}{A_0}
+w_3\frac{\partial^2\dot{\psi}}{a^2A_0}+w_6\frac{\partial^2\psi}{a^2}\right)\alpha
\right.\notag\\
\hspace{-0.7cm}
&&\left.~~~
-w_3\frac{(\partial\da)^2}{4a^2A_0^2}+w_5\frac{\da^2}{A_0^2}
+\left\{w_3\dot{\psi}-(w_2-A_0w_6)\psi\right\}\frac{\partial^2\da}{2a^2A_0^2}
-w_3\frac{(\partial\dot{\psi})^2}{4a^2A_0^2}+w_7\frac{(\partial\psi)^2}{2a^2}
\right]\,,\label{LGP}\\
\hspace{-0.7cm}
{\cal L}^{\rm flat}_{3}
&=&a^3 \left[ \left( \rho_m+P_m \right)v\frac{\partial^2 \chi}
{a^2}-v\dot{\delta \rho}_m-3H (1+c_m^2) v\delta \rho_m 
-\frac{1}{2} (\rho_m+P_m) \frac{(\partial v)^2}{a^2}
-\frac{c_m^2}{2 (\rho_m+P_m)} \delta \rho_m^2 
-\alpha \delta \rho_m \right]\,,
\label{L3}\\
\hspace{-0.7cm}
{\cal L}_{\zeta}
&=& a^3\left[
\left\{
3D_6\dot{\dphi}-3D_7\dphi-3w_1\alpha+\frac{3w_2}{A_0}\da-3(\rho_m+P_m)v
+\frac{2}{a^2}(q_t\pa^2\chi+\alpha_3\pa^2\psi)\right\}\dot{\zeta}-3q_t\dot{\zeta}^2
\right.\notag\\
\hspace{-0.7cm}
&&\left.~~~
-\left\{B_1\dphi+2(q_t-2A_0\alpha_3)\alpha+2\alpha_3\da\right\}\frac{\pa^2\zeta}{a^2}
+q_t c_t^2\frac{(\pa\zeta)^2}{a^2}
\right]\,,
\label{Lze}\\
\hspace{-0.7cm}
{\cal L}_{E}
&=& a^3\Bigg[
2q_t\ddot{\zeta}+2B_2\dot{\zeta}
-D_6\ddot{\dphi}-B_3\dot{\dphi}+B_4\dphi
+w_1\dot{\alpha}+(\dot{w_1}+3Hw_1)\alpha
-\frac{w_2}{A_0}\dot{\da}+B_5\da
\notag\\
\hspace{-0.7cm}
&&~~~
+(\rho_m+P_m)(\dot{v}-3Hc_m^2v)
\Bigg]\,\pa^2 E\,, 
\label{LE}
\ea
where $q_t$, $c_t^2$, $\alpha_3$ are given by Eqs.~(\ref{qt}), 
(\ref{ct}), (\ref{al3}), respectively, and the explicit forms of 
coefficients $D_{1,...,10}, w_{1,...,8}$ are shown in Appendix \ref{coeff}.  
The effect of intrinsic vector modes on scalar perturbations 
appears through the quantity 
\be
w_3=-2A_0^2\,q_v\,.
\ee
The coefficients $B_{1,...,5}$ in Eqs.~(\ref{Lze})-(\ref{LE}) 
can be expressed by using other coefficients, as
\ba
&&
B_1=\frac{2}{\tp}\left[\dot{q_t}+(1-c_t^2)Hq_t-A_0(\dot{\alpha_3}+H\alpha_3)\right]\,,
\qquad 
B_2=\dot{q_t}+3Hq_t\,,
\qquad 
B_3=\dot{D_6}+3HD_6-D_7\,,
\notag\\
&&
B_4=\dot{D_7}+3HD_7\,,
\qquad
B_5=-\frac{1}{A_0} \left[ \dot{w_2}+3Hw_2
+\dot{A_0}(w_6-4H\alpha_3) \right]\,.
\label{Bi}
\ea
The first three Lagrangians 
${\cal L}^{\rm flat}_{1},{\cal L}^{\rm flat}_{2}, {\cal L}^{\rm flat}_{3}$ in Eq.~(\ref{Ss}) 
are equivalent to those derived for the flat gauge in Ref.~\cite{Kasedark}. 
The other two Lagrangians ${\cal L}_{\zeta}$ and 
${\cal L}_{E}$ arise from metric perturbations 
$\zeta$ and $E$, respectively.

Since the perturbations $\alpha,\chi,\da,v,E$ do not possess their kinetic terms in the second-order action (\ref{Ss}), they correspond to nondynamical variables.  
Varying the cation (\ref{Ss}) with respect to 
$\alpha,\chi,\da,v,E$, we obtain 
their equations of motion in Fourier space, as 
\ba
\hspace{-.5cm}
\E_{\alpha} &\equiv&
D_4\dot{\dphi}-3w_1\dot{\zeta}
+D_5\dphi+2w_4\alpha+w_8\frac{\da}{A_0}
+\frac{k^2}{a^2}\left[ 2(q_t-2A_0\alpha_3)\zeta+w_6\psi-w_1\chi-D_6\dphi
-{\cal Y}+a^2w_1\dot{E} \right]-\delta \rho_m \nonumber \\
& &=0\,,
\label{eqalpha}\\
\hspace{-.5cm}
\E_{\chi} &\equiv&
D_6\dot{\dphi}-2q_t\dot{\zeta}-D_7\dphi-w_1\alpha
-\left( \rho_m+P_m \right)v
+w_2\frac{\da}{A_0}=0\,,
\label{eqchi}\\
\hspace{-.5cm}
\E_{\delta A} &\equiv&
D_8\dot{\dphi}+3w_2\frac{\dot{\zeta}}{A_0}
+D_9\dphi+w_8\frac{\alpha}{A_0}+2w_5\frac{\da}{A_0^2}
+\frac{k^2}{a^2}\frac{1}{A_0} 
\left( 
2A_0\alpha_3\zeta+w_2\chi-\frac{A_0w_6-w_2}{2A_0}\psi
+\frac{1}{2}{\cal Y}-a^2w_2\dot{E}\right) \nonumber \\
& &=0\,,
\label{eqdA}\\
\hspace{-.5cm}
\E_{v} &\equiv&
\dot{\delta \rho}_m+3H \left( 1+c_m^2 \right)
\delta \rho_m
+3(\rho_m+P_m)\dot{\zeta}
+\frac{k^2}{a^2} \left( \rho_m+P_m \right) 
\left( v+\chi-a^2\dot{E} \right)=0\,,
\label{eqdrho} \\
\hspace{-.5cm}
\E_{E} &\equiv&
2q_t\ddot{\zeta}+2B_2\dot{\zeta}
-D_6\ddot{\dphi}-B_3\dot{\dphi}+B_4\dphi
+w_1\dot{\alpha}+(\dot{w_1}+3Hw_1)\alpha
-\frac{w_2}{A_0}\dot{\da}+B_5\da
+(\rho_m+P_m)(\dot{v}-3Hc_m^2v) \nonumber \\
& &
=0\,,
\label{eqE}
\ea
where $k$ is a comoving wavenumber, and 
\be
{\cal Y} \equiv -\frac{w_3}{A_0} 
\left( \dot{\psi}+\delta A-2 A_0\alpha \right)\,.
\label{Ydef}
\ee
To simplify Eq.~(\ref{eqdA}), we used
Eq.~(\ref{Bi}) and the following relation
\be
w_2+A_0 w_6=4H A_0 \alpha_3\,.
\label{property1}
\ee

Variations of the action (\ref{Ss}) with respect to the remaining perturbations
$\psi,\dphi,\delrho,\zeta$ lead to 
\ba
\E_{\psi} &\equiv&
\dot{\cal Y}+\left( H -\frac{\dot{A}_0}{A_0} \right){\cal Y}
+4A_0\alpha_3\dot{\zeta}
-\frac{1}{A_0} \left[ (2w_6 \alpha+2w_7 \psi
-2D_{10} \delta \phi) 
A_0^2+(w_2-w_6 A_0 )\delta A \right]=0\,,\label{calYeq}\\
\E_{\delta \phi} &\equiv&
\dot{\cal Z}+3H {\cal Z}+3D_7\dot{\zeta}-2D_3 \delta \phi-D_5 \alpha-D_9 \delta A
-\frac{k^2}{a^2} \left( 2D_2 \delta \phi -D_6 \alpha 
-D_7 \chi-D_{10}\psi+B_1\zeta-a^2B_4E
 \right) \nonumber \\
& & =0\,,\label{calZeq}\\
\E_{\delta \rho_m} &\equiv&
\dot{v}-3Hc_m^2 v-\frac{c_m^2}{\rho_m+P_m} \delta \rho_m 
-\alpha=0\,,
\label{veq}\\
\E_{\zeta} &\equiv&
\dot{\cal W} +3H{\cal W}+(\rho_m+P_m)(\dot{v}-3Hc_m^2v)
+\frac{k^2}{3a^2}\left[2(q_t-2A_0\alpha_3)\alpha+2q_tc_t^2\zeta
+B_1\dphi+2\alpha_3\da\right]=0\,,
\label{calWeq}
\ea
where 
\ba
{\cal Z} &\equiv& 2D_1 \dot{\delta \phi}+3D_6\dot{\zeta}
+D_4 \alpha+D_8 \delta A
+\frac{k^2}{a^2}\left[D_6\chi 
-a^2(D_6\dot{E}+D_7E)
\right]\,, \label{Zdef}\\
{\cal W} &\equiv& 2q_t\dot{\zeta}-D_6\dot{\dphi}+D_7\dphi+w_1\alpha-
\frac{w_2}{A_0}\da+\frac{2k^2}{3a^2}(q_t\chi+\alpha_3\psi
-q_t a^2 \dot{E})\,. 
\label{Wdef}
\ea
The second-order time derivatives 
$\ddot{\zeta}$ and $\ddot{\delta \phi}$ can be eliminated by 
combining Eq.~(\ref{eqE}) with (\ref{calWeq}). 
On using Eqs.~(\ref{Bi}) and (\ref{property1}) as well, 
we obtain
\be
q_t \left( \alpha+\dot{\chi}
+c_t^2 \zeta+H\chi-a^2 \ddot{E}-3a^2 H \dot{E}  
\right)+  \dot{q}_t \left( \chi-a^2 \dot{E} \right)
+\frac{B_1}{2} \left( \delta \phi-\frac{\dot{\phi}}{A_0} 
\psi \right)
-\left[ H (c_t^2-1)q_t-\dot{q}_t \right]
\frac{\psi}{A_0}-\frac{A_0 \alpha_3{\cal Y}}{w_3}=0\,. 
\label{Bianchi}
\ee
The second-order action (\ref{Ss}) and the linear perturbation 
Eqs.~(\ref{eqalpha})-(\ref{eqE}), (\ref{calYeq})-(\ref{calWeq}), and (\ref{Bianchi}) are valid for arbitrary gauges and hence they are written in gauge-ready forms. 

Let us confirm the consistency of scalar perturbation 
equations of motion derived above. 
In doing so, we employ the following relations:
\ba
& &
D_1 \dot{\phi}^2=-3H^2 q_t-3H (w_1-w_2)
+w_4+w_5+w_8\,,
\label{spere1}\\
& &
D_4 \dot{\phi}=3H w_1-2w_4-w_8\,,\\
& &
D_8 \dot{\phi}A_0=-3H w_2-2w_5-w_8\,,\\
\label{spere3}
& &
D_6 \dot{\phi}=w_1-w_2+2Hq_t\,,\label{spere4}\\
&&
A_0\tp D_{10}=A_0^2w_7+\dot{A_0}w_6-2(A_0\dot{H}+\dot{A_0}H)\alpha_3\,,\\
&&
2A_0\tp(\tp D_2+D_7)=A_0^3w_7+2A_0^2(H\dot{\alpha_3}-\dot{H}\alpha_3+H^2\alpha_3)
-2\dot{A_0}w_2
\notag\\
&&
\hspace{3cm}
+A_0\left[2H^2q_t(c_t^2-2)+H(2\dot{A_0}\alpha_3+w_2-w_1)
-2\tp\dot{D_6}+\dot{w_1}-\dot{w_2}-\rho_m-P_m\right]\,, 
\label{spere6}\\
&&
2\tp D_{3}=\frac{1}{a^3}\frac{d}{dt}(a^3D_5)-\frac{3H}{a^3}\frac{d}{dt}(a^3D_7)
+\frac{1}{a^3A_0}\frac{d}{dt}(a^3A_0^2D_9)\,,
\label{spere7}
\ea
as well as their time derivatives. Using these properties 
and the background Eqs.~(\ref{back2}), (\ref{back3}), (\ref{coneq}), 
(\ref{back5}) and (\ref{property1}), it follows that there are
two particular relations among the perturbation equations:
\ba
&&
\frac{1}{a^3}\frac{d}{dt}\left(a^3\E_{\alpha}\right)
-3H\E_{\zeta}
+\frac{1}{a^3A_0}\frac{d}{dt}(a^3A_0^2\E_{\da})
-\tp\,\E_{\dphi}
-\frac{k^2}{a^2}(\E_{\chi}+A_0\E_{\psi})
+3H(\rho_m+P_m) \E_{\delta \rho_m}+\E_v=0\,,
\label{Bianchi1}\\
&&
\E_{E}-\frac{1}{a^3}\frac{d}{dt}\left(a^3\E_{\chi}\right)
=0\,, 
\label{Bianchi2}
\ea
which correspond to the temporal and spatial components 
of the Bianchi identity, respectively. 
Thus, we have confirmed the 
consistency of Eqs.~(\ref{eqalpha})-(\ref{eqE}) and 
(\ref{calYeq})-(\ref{calWeq}) with the Bianchi identity.

\subsection{Gauge transformations and the choice of gauges}
\label{gaugesec}

Now, we discuss the issue of gauge transformations, 
gauge-invariant variables, and gauge fixings.
We consider the scalar gauge transformation from the coordinate 
$x^{\mu}=(t,x^i)$ to another coordinate 
$\tilde{x}^{\mu}=(\tilde{t},\tilde{x}^i)$, as 
\be
\tilde{t}=t+\xi^{0}\,,\qquad 
\tilde{x}^{i}=x^{i}+\delta^{ij} \partial_{j} \xi\,,
\label{gaugetra}
\ee
where $\xi^{0}$ and $\xi$ determine the time slicing and 
spatial threading, respectively. 
The four scalar metric perturbations 
$\alpha, \chi, \zeta, E$ transform as 
\ba
\tilde{\alpha}=\alpha-\dot{\xi}^{0}\,,\qquad 
\tilde{\chi}=\chi+\xi^{0}-a^2 \dot{\xi}\,,\qquad 
\tilde{\zeta}=\zeta-H \xi^{0}\,,\qquad 
\tilde{E}=E-\xi\,.
\ea
The transformations of scalar-field perturbation $\delta \phi$ 
and matter density perturbation $\delta \rho_m$ are given by 
\be
\widetilde{\delta \phi}=\delta \phi-\dot{\phi}\,\xi^{0}\,,
\qquad 
\widetilde{\delta \rho_m}=\delta \rho_m-\dot{\rho}_m\,\xi^{0}\,.
\label{delrho}
\ee
For the vector field $A_{\mu}$, we use the property that the 
scalar product $A_{\mu}dx^{\mu}$ is invariant 
under the gauge transformation. 
This leads to the following relations 
\be
\widetilde{\delta A}=
\delta A-A_0 \dot{\xi}^{0}+\dot{A}_0 \xi^{0}\,,
\qquad \tilde{\psi}=\psi-A_0 \xi^{0}\,.
\ee
The velocity potential $v$ transforms as 
\be
\tilde{v}=v-\xi^{0}\,.
\label{v}
\ee

We can construct several perturbed quantities invariant 
under the transformation (\ref{gaugetra}).
The gauge-invariant Bardeen gravitational 
potentials are given by \cite{Bardeen}
\be
\Psi=\alpha+\dot{\chi}-\frac{d}{dt} 
\left( a^2 \dot{E} \right)\,,\qquad 
\Phi=\zeta+H \chi-a^2 H\dot{E} \,,
\label{PsiPhi}
\ee
which are commonly used for the study of cosmic growth 
history in the presence of dark energy. 
There are also the following gauge-invariant quantities:
\ba
& &
\delta \phi_{\rm f}=\delta \phi
-\frac{\dot{\phi}}{H}\zeta\,,
\qquad \delta \phi_{\rm v}=
\delta \phi-\frac{\dot{\phi}}{A_0}\psi\,,
\qquad 
\delta \phi_{\rm N}=\delta \phi+\dot{\phi}\chi
-a^2 \dot{\phi} \dot{E}\,,
\label{delphif}\\
& &
\psi_{\rm f}=\psi
-\frac{A_0}{H}\zeta\,,\qquad 
\psi_{\rm u}= \psi-\frac{A_0}{\dot{\phi}}\delta \phi\,,
\qquad 
\psi_{\rm N}=\psi+A_0 \chi-a^2A_0 \dot{E}\,,
\label{psif}\\
& &
\delta \rho_{\rm f}=\delta \rho_m -\frac{\dot{\rho}_m}{H}\zeta\,,\qquad 
\delta \rho_{\rm u}=\delta \rho_m -\frac{\dot{\rho}_m}
{\dot{\phi}}\delta \phi\,,\qquad  
\delta \rho_{\rm v}=\delta \rho_m -\frac{\dot{\rho}_m}
{A_0}\psi\,,\qquad  
\delta \rho_{\rm N}=
\delta \rho_m+\dot{\rho}_m \chi-a^2\dot{\rho}_m \dot{E}\,,
\label{rhof}\\
& &
\delta_m=\frac{\delta \rho_m}{\rho_m}
+3H \left( 1+\frac{P_m}{\rho_m} \right)v\,,
\ea
where $\delta \phi_{\rm f}$ is called the 
Mukhanov-Sasaki variable \cite{Mukhanov,Sasaki}. 

In the context of inflationary cosmology, it is convenient to 
introduce the following gauge-invariant curvature 
perturbations \cite{Lukash,Lyth}:
\be
{\cal R}_{\phi}=\zeta-\frac{H}{\dot{\phi}} \delta \phi
\,,\qquad 
{\cal R}_{\psi}=\zeta-\frac{H}{A_0} \psi\,.
\label{Rpc}
\ee
We define the time derivative of an adiabatic field $\sigma$ 
representing the velocity along the 
background trajectory, as \cite{Gordon}
\be
\dot{\sigma}=(\cos \theta) \dot{\phi}+(\sin \theta) M A_0\,,
\label{dotsigma}
\ee
where $M$ is a constant having a dimension of mass, and 
\be
\cos \theta=\frac{\dot{\phi}}
{\sqrt{\dot{\phi}^2+M^2 A_0^2}}\,,\qquad 
\sin \theta=\frac{M A_0}{\sqrt{\dot{\phi}^2+M^2 A_0^2}}\,.
\ee
The adiabatic field perturbation $\delta \sigma$ and the entropy 
perturbation $\delta s$ orthogonal to the background 
trajectory are 
defined, respectively, by 
\ba
\delta \sigma &=& (\cos \theta) \delta \phi+ (\sin \theta) M\psi\,,
\label{delsig}\\
\delta s &=&  (\cos \theta) M\psi- (\sin \theta) \delta \phi\,,
\label{dels}
\ea
where $\delta s$ is gauge-invariant by construction. 
We also introduce the total gauge-invariant 
curvature perturbation incorporating 
both $\delta \phi$ and $\psi$, as 
\ba
{\cal R} &=&  (\cos^2 \theta){\cal R}_{\phi}
+ (\sin^2 \theta){\cal R}_{\psi} \nonumber \\
&=&
\zeta-\frac{H(\dot{\phi}\,\delta \phi+M^2A_0 \psi)}
{\dot{\phi}^2+M^2A_0^2}\,.
\label{calR}
\ea
In terms of the adiabatic field $\sigma$ and its perturbation $\delta \sigma$, 
Eq.~(\ref{calR}) can be expressed as 
\be
{\cal R}=\zeta-\frac{H \delta \sigma}{\dot{\sigma}}\,.
\label{calR2}
\ee
For the background field trajectory 
satisfying $\dot{\theta} \neq 0$, 
the entropy perturbation $\delta s$ generally works as a source 
term for the adiabatic perturbation $\delta \sigma$ \cite{Gordon}. 
Hence the evolution of ${\cal R}$ is known by studying how 
$\delta s$ and $\delta \sigma$ evolve 
in time \cite{Wands,TPB}. 

On using the gauge-invariant variables (\ref{PsiPhi}) 
and (\ref{Rpc}), we can write Eq.~(\ref{Bianchi}) 
in the following simple form
\be
\Psi+\left( 1+\alpha_{\rm M} \right) \Phi
+\left( c_t^2-1-\alpha_{\rm M} \right){\cal R}_{\phi}
+\frac{A_0}{Hq_t} \left( \dot{\alpha}_3+H\alpha_3 
\right) \left( {\cal R}_{\phi}-{\cal R}_{\psi} \right)
-\frac{A_0\alpha_3}{q_t w_3}{\cal Y}=0\,,
\label{anire}
\ee
where 
\be
\alpha_{\rm M} \equiv \frac{\dot{q}_t}{Hq_t}\,.
\label{alM}
\ee
We note that the perturbation ${\cal Y}$ 
defined by  Eq.~(\ref{Ydef}) is also gauge-invariant.

Let us consider theories satisfying the condition 
\be
\alpha_3 =
-2A_0 f_{4,X_3}-HA_0^2 f_{5,X_3}
+\dot{\phi}f_{5,\phi}
=0\,. \label{al30}
\ee
Then, Eq.~(\ref{anire}) reduces to 
\be
\Psi+\left( 1+\alpha_{\rm M} \right) \Phi
+\left( c_t^2-1-\alpha_{\rm M} \right){\cal R}_{\phi}=0 
\qquad ({\rm for}~\,\alpha_3=0).
\label{anire2}
\ee
The condition (\ref{al30}) is satisfied not only for
Horndeski theories but also for
SVT theories with the couplings:
\be
f_4=f_4(\phi)\,,\qquad f_5={\rm constant}\,.
\label{coup}
\ee
In such cases, the time variation of $q_t$ (i.e., 
$\alpha_{\rm M} \neq 0$) and the deviation of 
$c_t^2$ from 1 give rise to the gravitational slip 
($-\Psi \neq \Phi$). 
For SVT theories with the couplings $f_4=f_4(X_3)$ and 
$f_5=f_5(\phi, X_3)$, the last two terms in 
Eq.~(\ref{anire}) also work as additional 
anisotropic stresses.

Under the transformation (\ref{gaugetra}), there are residual 
gauge DOFs for fixing $\xi^0$ and $\xi$. 
Several gauge conditions commonly 
used in the literature are 
\ba
& &
\zeta=0\,,\qquad E=0\,,\qquad\, \mathrm{(Flat~gauge)},
\label{flat}\\
& &
\delta \phi=0\,,\quad ~\,E=0\,,\qquad\, \mathrm{(Unitary~gauge)},\label{gauge3}\\
& &
\psi=0\,,\quad ~~~E=0\,,\qquad \mathrm{(Uniform~ vector~gauge)},\label{gaugev}\\
& &
\chi=0\,,\qquad E=0\,,\qquad \mathrm{(Newtonian~gauge)},\label{gauge1}\\
& &
\alpha=0\,,\qquad \chi=0\,,\qquad\, \mathrm{(Synchronous~gauge)}.\
\ea
Apart from the synchronous gauge in which $\xi^0$ is not 
unambiguously fixed, the other gauges (\ref{flat})-(\ref{gauge1}) 
completely fix $\xi^0$ and $\xi$. 

For the flat gauge, the dynamical DOFs correspond to 
the perturbations $\delta \phi_{\rm f}, \psi_{\rm f}$, 
and $\delta \rho_{\rm f}$. 
In Refs.~\cite{HKT18,Kasedark}, 
the second-order action of these dynamical fields  
was derived by choosing the flat gauge.
This gauge choice is valid in the expanding Universe 
($H>0$), but as we see in Sec.~\ref{appsec}, it is not 
suitable for describing the evolution of perturbations 
in bouncing cosmologies. 
This is not generally the case for gauges in which the 
perturbation $\zeta$ does not vanish, 
e.g., (\ref{gauge3})-(\ref{gauge1}). 

If we apply SVT theories to dark energy and choose the 
flat gauge, the contributions of tensor, vector, and 
scalar perturbations in observables associated with 
the cosmic growth are not transparent \cite{Kasedark}.
In Sec.~\ref{conmodelsec}, we show that the separation between 
tensor, vector, and scalar modes becomes clear by 
choosing gauges in which $\zeta$ does not vanish, e.g., 
the unitary gauge.

Thus, in our gauge-ready formulation, we can choose most 
suitable and convenient gauges depending on the problem 
at hand. While the underlying physics is not affected 
by the choice of different gauges,
it makes sense to choose most appropriate gauges in which 
the physical meaning and interpretation of results 
become transparent.

\section{Stability conditions in unitary and Newtonian gauges}
\label{stasec}

In this section, we derive conditions for the absence of 
ghost and Laplacian instabilities of scalar perturbations 
in the small-scale limit by choosing the unitary, uniform 
vector, and Newtonian gauges. 
In Ref.~\cite{Kasedark}, the similar analysis was performed 
in the flat gauge, but this gauge choice is not 
suitable for studying the evolution of curvature 
perturbations in the bouncing cosmology (as we will 
discuss in Sec.~\ref{appsec}).
This problem can be circumvented by choosing other 
suitable gauges discussed in this section.

\subsection{Unitary gauge}

Let us first consider the unitary gauge characterized by 
Eq.~(\ref{gauge3}).
In this case, the dynamical perturbations 
correspond to $\psi_{\rm u}=\psi$, 
${\cal R}_{\phi}=\zeta$, and 
$\delta \rho_{\rm u}=\delta \rho_m$, which are represented 
by the matrix
\be
\vec{\mathcal{X}}^{t}=\left(\psi_{\rm u}, {\cal R}_{\phi}, 
\delta \rho_{\rm u}/k \right) \,.
\ee
{}From Eqs.~(\ref{eqalpha})-(\ref{eqdrho}), 
the nondynamical perturbations $\alpha, \chi, \delta A, v$ can 
be expressed in terms of $\psi_{\rm u}, {\cal R}_{\phi}, 
\delta \rho_{\rm u}$ and 
their time derivatives. 
Substituting them into Eq.~(\ref{Ss}) and integrating it 
by parts, the second-order scalar action in Fourier 
space reduces to 
\be
{\cal S}_s^{(2)}=\int dt d^3x\,a^{3}\left( 
\dot{\vec{\mathcal{X}}}^{t}{\bm K}\dot{\vec{\mathcal{X}}}
-\frac{k^2}{a^2}\vec{\mathcal{X}}^{t}{\bm G}\vec{\mathcal{X}}
-\vec{\mathcal{X}}^{t}{\bm M}\vec{\mathcal{X}}
-\vec{\mathcal{X}}^{t}{\bm B}\dot{\vec{\mathcal{X}}}
\right)\,,
\label{Ss2}
\ee
where ${\bm K}$, ${\bm G}$, ${\bm M}$, ${\bm B}$ 
are $3 \times 3$ matrices. 
The leading-order contributions to ${\bm M}$ and ${\bm B}$ 
correspond to the order of $k^0$.
In the small-scale limit,  the nonvanishing matrix 
components of ${\bm K}$ and ${\bm G}$ are given by 
\ba
& &
K_{11}=\frac{w_1^2 w_5+w_2^2w_4+w_1 w_2 w_8}
{A_0^2 (w_1-2w_2)^2}\,,\qquad 
K_{22}=q_t \left[ 3+\frac{4q_t (w_4+4w_5+2w_8)}
{(w_1-2w_2)^2} \right]\,, \nonumber \\
& &
K_{12}=K_{21}
=\frac{q_t \left[ w_1 (4w_5+w_8)+2w_2 (w_4+w_8)
\right]}{A_0 (w_1-2w_2)^2}\,,\qquad
K_{33}=\frac{a^2}{2(\rho_m+P_m)}\,,
\label{Kcom}
\ea
and 
\ba
& &
G_{11}=\frac{\alpha_2}{2}+\frac{2E_1^2}{q_v}
-\frac{w_2^2 (\rho_m+P_m)}{2A_0^2 (w_1-2w_2)^2}
+\frac{1}{a} \frac{d}{dt} \left( aE_1 \right)
\,,\qquad
G_{22}=-q_t c_t^2+\frac{2E_3^2}{q_v}
-\frac{2q_t^2 (\rho_m+P_m)}{(w_1-2w_2)^2}
+\frac{1}{a} \frac{d}{dt} \left( aE_2 \right)
\,, \nonumber \\
& &
G_{12}=G_{21}
=\frac{2E_1E_3}{q_v}-\frac{w_2q_t (\rho_m+P_m)}
{A_0 (w_1-2w_2)^2}+\frac{1}{a} \frac{d}{dt} \left( aE_3 \right)\,,\qquad
G_{33}=\frac{c_m^2 a^2}{2(\rho_m+P_m)}\,,
\label{Gcom}
\ea
where we used the relation $\alpha_2=-w_7$, and 
\be
E_1=\frac{w_6}{4A_0}-\frac{w_1w_2}
{4A_0^2(w_1-2w_2)}\,,\qquad 
E_2=-\frac{2q_t^2}{w_1-2w_2}\,,\qquad 
E_3=-\frac{w_2+A_0 w_6}{4H A_0}
-\frac{q_t w_2}{A_0 (w_1-2w_2)}\,.
\label{E123}
\ee
The last time derivatives in 
$G_{11}, G_{22}, G_{12}, G_{21}$ 
arise from partial integrations of the terms containing 
$k^2/a^2$ in $\vec{\mathcal{X}}^{t}{\bm B}\dot{\vec{\mathcal{X}}}$.
The matter perfect fluid is decoupled from other fields 
$\psi_{\rm u}$ and ${\cal R}_{\phi}$, 
so that the ghost and Laplacian instabilities are absent 
for $\rho_m+P_m>0$ and $c_m^2>0$. 
The quantities $K_{11}$ and $G_{11}$ are identical to 
those derived for the flat gauge in Ref.~\cite{Kasedark}, 
but $K_{22}, K_{12}, G_{22}, G_{12}$ are different 
by reflecting the fact that ${\cal R}_{\phi}$ corresponds
to the dynamical DOF in the unitary gauge (unlike 
$\delta \phi_{\rm f}$ in the flat gauge).

The conditions for the absence of scalar ghosts are given by 
$K_{11}>0$ or $K_{22}>0$, and 
\be
q_s \equiv K_{11}K_{22}-K_{12}^2>0\,.
\label{qsdef}
\ee
In the unitary gauge, the quantity $q_s$ reduces to 
\be
q_s^{\rm (u)}=\frac{q_t [q_t(4w_4w_5-w_8^2)+
3(w_1^2 w_5+w_1w_2w_8+w_2^2 w_4)]}
{A_0^2 (w_1-2w_2)^2}\,.
\ee
On using the properties (\ref{spere1})-(\ref{spere3}), 
the determinant ${\cal D}$ defined by Eq.~(\ref{calD})
is expressed in the form 
\be
{\cal D}=\frac{2}{\dot{\phi}^2} 
\left[ q_t(4w_4w_5-w_8^2)+
3(w_1^2 w_5+w_1w_2w_8+w_2^2 w_4) \right]\,.
\ee
Then, $q_s^{\rm (u)}$ is proportional to ${\cal D}$, as 
\be
q_s^{\rm (u)}=\frac{\dot{\phi}^2 q_t}
{2A_0^2 (w_1-2w_2)^2} {\cal D}\,.
\label{qsu}
\ee
Since $q_s^{\rm (u)}>0$ and $q_t>0$ for the absence 
of scalar and tensor ghosts, the determinant associated with 
the closed-form background equations of motion 
needs to be in the range 
\be
{\cal D}>0\,.
\label{noghocon}
\ee
In the flat gauge chosen in Refs.~\cite{Heisenberg18,Kasedark}, 
the quantity (\ref{qsdef}) is given by 
\be
q_s^{\rm (f)}=\frac{H^2 q_t}
{2A_0^2 (w_1-2w_2)^2} {\cal D}
=\frac{H^2}{\dot{\phi}^2}q_s^{\rm (u)}\,,
\label{qsre}
\ee
which is different from $q_s^{\rm (u)}$ only by 
an overall factor $H^2/\dot{\phi}^2$.

Taking the small-scale limit in Eq.~(\ref{Ss2}), the 
dispersion relation yields
${\rm det} (c_s^2 {\bm K}-{\bm G})=0$, where $c_s$ is 
the propagation speed of scalar perturbations. 
One of the solutions is the matter propagation speed 
squared $c_m^2$, while the other two solutions are 
\be
c_{s1}^2=\frac{{\cal F}_s}{2q_s} \left[ 1
+\sqrt{1-\frac{4q_s{\cal G}_s}{{\cal F}_s^2}} 
\right]\,,\qquad 
c_{s2}^2=\frac{{\cal F}_s}{2q_s} \left[ 1
-\sqrt{1-\frac{4q_s{\cal G}_s}{{\cal F}_s^2}} 
\right]\,,
\label{cs}
\ee
where 
\be
{\cal F}_s \equiv K_{11}G_{22}+K_{22}G_{11}
-2K_{12}G_{12}\,,\qquad 
{\cal G}_s \equiv G_{11}G_{22}-G_{12}^2\,.
\label{FGs}
\ee
To avoid small-scale Laplacian instabilities, we require 
the two conditions $c_{s1}^2>0$ and $c_{s2}^2>0$. 

In the flat gauge, the matrix components of ${\bm K}$ and 
${\bm G}$ contain the terms 
$D_1, D_2, D_4, D_6, D_7, D_8, D_{10}$ besides $w_i$, 
see Eqs.~(5.22) and (5.23) of Ref.~\cite{Kasedark}. 
On using  Eqs.~(\ref{back2}), (\ref{property1}), 
and (\ref{spere1})-(\ref{spere6}), we find that 
the quantities ${\cal F}_s$ and ${\cal G}_s$ in the unitary 
gauge are related to those in the flat gauge, as
\be
{\cal F}_s^{{\rm (u)}}=\frac{\dot{\phi}^2}{H^2}
{\cal F}_s^{{\rm (f)}}\,,\qquad 
{\cal G}_s^{{\rm (u)}}=\frac{\dot{\phi}^2}{H^2}
{\cal G}_s^{{\rm (f)}}\,.
\ee
Since there is also the correspondence (\ref{qsre}), 
it follows that the scalar propagation speeds are 
the same in both unitary and flat gauges.

\subsection{Uniform vector gauge}
\label{univec}

Let us consider the uniform vector gauge characterized by 
Eq.~(\ref{gaugev}).
In this case, the dynamical DOFs are given by the matrix 
\be
\vec{\mathcal{X}}^{t}=\left(\dphi_{\rm v}, {\cal R}_{\psi}, 
\delta \rho_{\rm v}/k \right)\,.
\ee
On using Eqs.~(\ref{eqalpha})-(\ref{eqdrho}) for $\alpha,\chi,\da,v$ 
to eliminate these nondynamical variables in the action (\ref{Ss}) 
and taking the small-scale limit, 
the resulting action is of the form (\ref{Ss2}) with 
the same values of $K_{22}, K_{33}$ and $G_{22}, G_{33}$ 
as those given in Eqs.~(\ref{Kcom}) and (\ref{Gcom}).
The other nonvanishing matrix components of ${\bm K}$ 
and ${\bm G}$ are
\ba
& &
K_{11}=
D_1+\frac{D_6}{w_1-2w_2}\left[D_4+2A_0D_8+\frac{D_6(w_4+4w_5+2w_8)}{w_1-2w_2}\right]\,, \nonumber \\
& &
K_{12}=K_{21}
=-\frac{q_t}{w_1-2w_2}\left[D_4+2A_0D_8
+\frac{2D_6(w_4+4w_5+2w_8)}{w_1-2w_2}\right]\,,
\label{KcomV} \\
& &
G_{11}=
-D_2-\frac{1}{w_1-2w_2}\left[D_6D_7+\frac{w_2^2F_1}{A_0^2q_v}
-(\rho_m+P_m)F_1\right]+\frac{1}{a} \frac{d}{dt} 
\left( a F_1 \right)\,,\nonumber \\
& &
G_{12}=G_{21}=
-\frac{B_1}{2}-\frac{\alpha_3w_2F_2}{A_0q_tq_v}
+\frac{1}{w_1-2w_2}\left[q_tD_7-\frac{w_2^2F_2}{A_0^2q_v}+(\rho_m+P_m)F_2\right]
+\frac{1}{a} \frac{d}{dt} \left( a F_2 \right)\,,
\ea
where
\be
F_1=-\frac{D_6^2}{2(w_1-2w_2)}\,,\qquad
F_2=\frac{q_tD_6}{w_1-2w_2}\,.
\ee
On using Eqs.~(\ref{spere1})-(\ref{spere4}), the quantity 
$q_s=K_{11}K_{22}-K_{12}^2$ in the uniform vector gauge 
is expressed as 
\be
q_s^{\rm (v)}=\frac{q_t}{2(w_1-2w_2)^2}{\cal D}\,.
\ee
Then, under the conditions ${\cal D}>0$ and $q_t>0$,
the scalar ghost is absent again. 
Compared to $q_s$ in the flat and unitary  
gauges, the following relations hold 
\be
q_s^{\rm (v)}=\frac{A_0^2}{H^2}q_s^{\rm (f)}=\frac{A_0^2}{\tp^2}q_s^{\rm (u)}\,.
\ee
Similarly, the quantities ${\cal F}_s$ and ${\cal G}_s$ 
in the uniform vector gauge are related to those 
in other gauges, as 
\be
{\cal F}_s^{\rm (v)}=\frac{A_0^2}{H^2}{\cal F}_s^{\rm (f)}=\frac{A_0^2}{\tp^2}{\cal F}_s^{\rm (u)}\,,\qquad
{\cal G}_s^{\rm (v)}=\frac{A_0^2}{H^2}{\cal G}_s^{\rm (f)}=\frac{A_0^2}{\tp^2}{\cal G}_s^{\rm (u)}\,.
\ee
Hence the scalar propagation speed squares (\ref{cs}) 
are identical to those in the flat and unitary gauges.

\subsection{Newtonian gauge}
\label{Newsec}

We also compute quantities associated with the stability of 
scalar perturbations in the Newtonian gauge (\ref{gauge1}). 
For the dynamical DOFs, we consider the following combinations:
\be
\vec{\mathcal{X}}^{t}=\left({\cal R}_{\psi}, {\cal R}_{\phi}, 
\delta \rho_{\rm N}/k \right)\,,
\label{dynaNew}
\ee
where ${\cal R}_{\psi}=\zeta-H \psi_{\rm N}/A_0$ and 
${\cal R}_{\phi}=\zeta-H \delta \phi_{\rm N}/\dot{\phi}$.
On using Eqs.~(\ref{eqalpha}), (\ref{eqdA}), and (\ref{eqdrho}), 
we first eliminate the nondynamical DOFs $\alpha, \delta A, v$ 
from the second-order scalar action. Then, we express
the perturbations $\psi_{\rm N}, \zeta$ and their time derivatives 
$\dot{\psi}_{\rm N}, \dot{\zeta}$ 
in terms of ${\cal R}_{\psi}, {\cal R}_{\phi}$ and their time 
derivatives $\dot{{\cal R}}_{\psi}, \dot{{\cal R}}_{\phi}$.

After this procedure, the term proportional to 
$\dot{\delta \phi}_{\rm N}^2$ vanishes 
from the second-order scalar action in the small-scale limit, 
so the perturbation $\delta \phi_{\rm N}$ behaves as a nondynamical DOF. 
After integrating terms containing 
$\dot{\delta \phi_{\rm N}}$ by parts, the contributions to the 
second-order scalar action arising from 
$\delta \phi_{\rm N}$ consist of the term 
$\delta \phi_{\rm N}^2$ and the products of 
$\delta \phi_{\rm N}$ and other dynamical perturbations
(say, $\delta \phi_{\rm N} \dot{\cal R}_{\phi}$). 
Varying this action with respect to 
$\delta \phi_{\rm N}$, we can express $\delta \phi_{\rm N}$ 
in terms of ${\cal R}_{\psi}, {\cal R}_{\phi}, \delta \rho_{\rm N}$
and their time derivatives.
Substituting this relation into Eq.~(\ref{Ss}), 
we obtain the second-order scalar action of the form (\ref{Ss2}) 
with the dynamical perturbations given by Eq.~(\ref{dynaNew}).

Again, the matter perturbation $\delta \rho_{\rm N}$ is decoupled from 
other dynamical fields ${\cal R}_{\psi}$ and ${\cal R}_{\phi}$.
The matrix components $K_{11}, K_{22}, K_{12}$ are not the same as 
those in the unitary or uniform vector gauges, but the combination 
$q_s=K_{11}K_{22}-K_{12}^2$ is related to each 
other among different gauges up to positive overall factors. 
Under the choice of the dynamical variables (\ref{dynaNew}), 
the quantity $q_s$ in the Newtonian gauge reads
\be
q_s^{\rm (N)}=\frac{\dot{\phi}^2 q_t}
{2H^2 (w_1-2w_2)^2}{\cal D}
=\frac{A_0^2}{H^2}q_s^{\rm (u)}
=\frac{\dot{\phi}^2}{H^2}q_s^{\rm (v)}
=\frac{A_0^2 \dot{\phi}^2}{H^4}q_s^{\rm (f)}\,,
\label{qsN}
\ee
and hence the scalar ghost is absent for 
${\cal D}>0$ and $q_t>0$.

In the Newtonian gauge, the quantities ${\cal F}_s$ 
and ${\cal G}_s$ are related to those in other gauges, as
\be
{\cal F}_s^{\rm (N)}=\frac{A_0^2}{H^2} 
{\cal F}_s^{\rm (u)}=
\frac{\dot{\phi}^2}{H^2}{\cal F}_s^{\rm (v)}=
\frac{A_0^2 \dot{\phi}^2}{H^4} {\cal F}_s^{\rm (f)}\,,
\qquad
{\cal G}_s^{\rm (N)}=\frac{A_0^2}{H^2} 
{\cal G}_s^{\rm (u)}=
\frac{\dot{\phi}^2}{H^2} {\cal G}_s^{\rm (v)}=
\frac{A_0^2 \dot{\phi}^2}{H^4} {\cal G}_s^{\rm (f)}\,,
\ee
which explicitly show that $c_{s1}^2$ 
and $c_{s2}^2$ are gauge-invariant quantities.

The difference of the quantities $q_s$, ${\cal F}_s$, ${\cal G}_s$ 
among several gauges simply comes from the choice 
of different dynamical perturbations. 
For instance, if we choose the perturbation 
$\psi_{\rm f}=-(A_0/H){\cal R}_{\psi}=\psi_{\rm N}-(A_0/H)\zeta$ 
besides ${\cal R}_{\phi}$ and $\delta \rho_{\rm N}/k$ as dynamical 
DOFs in the Newtonian gauge, it follows that 
$q_s, {\cal F}_s, {\cal G}_s$ coincide with those 
in the unitary gauge. 
The choice of the dynamical variables 
$\delta\phi_{\rm f}=-(\dot{\phi}/H){\cal R}_{\phi}
=\delta \phi_{\rm N}-(\dot{\phi}/H)\zeta$ besides 
${\cal R}_{\psi}$ and $\delta \rho_{\rm N}/k$ 
in the Newtonian gauge gives rise to the same values of 
$q_s, {\cal F}_s, {\cal G}_s$ as those in the 
uniform vector gauge.

\section{Application to nonsingular cosmology}
\label{appsec}

In this section, we apply the stability conditions derived 
in Sec.~\ref{stasec} to the nonsingular cosmology 
in which the scale factor is always in the region 
$a>0$. Our main interest is to discuss the possibility for 
realizing nonsingular bouncing/genesis solutions free from 
ghost and Laplacian instabilities.

\subsection{No-ghost condition at the bounce}

Let us first consider the bouncing cosmology in which 
the Universe transits from the collapse to the expansion.
Then, the Hubble parameter $H$ vanishes at the 
point of bounce. At $H=0$, the gauge-invariant 
perturbations $\delta \phi_{\rm f}, \psi_{\rm f}, 
\delta \rho_{\rm f}$ in Eqs.~(\ref{delphif})-(\ref{rhof}) 
are not well defined because their denominators vanish. 
For the flat gauge ($\zeta=0$), it looks as if such divergences 
can be circumvented, but the problem manifests in curvature 
perturbations defined by Eq.~(\ref{Rpc}). 
Since ${\cal R}_{\phi}=-H\delta \phi/\dot{\phi}$ and 
${\cal R}_{\psi}=-H\psi/A_0$ in the flat gauge, 
both ${\cal R}_{\phi}$ and ${\cal R}_{\psi}$ 
vanish at $H=0$.

Provided that $A_0 (w_1-2w_2) \neq 0$, 
the quantity $q_s^{{\rm (f)}}$ given by 
Eq.~(\ref{qsre}) is 0 at $H=0$. 
Then, it looks as if there were the strong coupling problem 
at the bounce, but this is an artifact of 
choosing the flat gauge in which ${\cal R}_{\phi}$ 
and ${\cal R}_{\psi}$ vanish at $H=0$. 
The gauge-invariant variables $\psi_{\rm u}$, 
${\cal R}_{\phi}$, and $\delta \rho_{\rm u}$, 
which reduce to $\psi$, $\zeta$, and $\delta \rho_m$ 
respectively in the unitary gauge ($\delta \phi=0$), are well 
defined except for $\dot{\phi}=0$. 
At the bounce, both ${\cal R}_{\phi}$ and 
${\cal R}_{\psi}$ reduce to $\zeta$. 
The right hand side of Eq.~(\ref{qsu}) does not contain 
terms proportional to $H^2$, so $q_s^{{\rm (u)}}$ 
does not vanish at $H=0$.  
The gauge-invariant perturbations 
$\delta \phi_{\rm v}, {\cal R}_{\psi}$, and $\delta \rho_{\rm v}$
in the uniform vector gauge are also well defined 
during the transition across the bounce, in which case 
$q_s^{(\rm v)}$ does not vanish at $H=0$.

The above discussion shows that the real physical problem 
arises only when the combination $q_t {\cal D}$, which
appears in $q_s$ for any gauge choice, crosses 0. 
Under the no-ghost condition of tensor perturbations 
($q_t>0$), this only occurs when ${\cal D}$ approaches 0. 
In the limit that ${\cal D} \to 0$, however, the background 
equations of motion exhibit the divergence. 
Thus, the crossing of $q_s=0$
can be avoided for the nonsingular background cosmology 
in which the  determinant is always in the range ${\cal D}>0$ 
and does not approach 0.

The above issue is closely related to the ``$\gamma$-crossing'' arising in 
Horndeski bouncing cosmologies with the violation of null energy 
conditions (NECs).
In Einstein gravity the quantity $\gamma$ is equivalent to 
$H$ \cite{Xue}, while in Horndeski gravity the mixing between the scalar kinetic 
energy and the metric (``braiding'' \cite{braiding}) leads to 
the difference from $\gamma$ from $H$ \cite{Ijjas}. 
For the bouncing solutions reducing to 
Einstein gravity both before and after the NEC violation, 
$\gamma$ crosses 0 during the transition from Einstein 
to Horndeski gravity. If we use the Hamiltonian and momentum 
constraints to eliminate the lapse perturbation $\alpha$ and 
the shear perturbation $\chi-a^2 \dot{E}$, then the quantity 
$\gamma$ appears in the denominators of these equations. 
The apparent divergence of lapse and shift at $\gamma=0$ 
is interpreted as a coordinate singularity 
that arises from a particular foliation of the spacetime. 
This coordinate singularity can be avoided by choosing a proper 
time slicing \cite{Ijjas}, so it does not correspond 
to a real, physical singularity.

\subsection{Possibility for realizing nonsingular cosmology}

There have been attempts for constructing bouncing/genesis 
cosmological solutions without the initial singularity. 
This requires the violation of NEC, 
which is not realized by conventional 
matter satisfying $\rho_m+P_m \geq 0$. 
Galileons and its generalizations \cite{Gali1,Gali2,Gali3} 
can be the candidates for violating the null energy condition. 
Indeed, generalized Galileons  
allow the existence of nonsingular bouncing solutions 
with neither ghost nor Laplacian instabilities around the 
bounce \cite{Qiu,Easson,Osipov,Anna,Vikman}. 
In the original genesis scenario and 
its variants \cite{Cremi1,Cremi2,Hint,Easson2,Elder,Pirt,Nishi}, 
it is possible to realize an initial super-accelerating stage without 
ghost and Laplacian instabilities.

Although generalized Galileons can give rise to nonsingular solutions 
stable near the bounce or during the super-accelerating stage, 
the stability of cosmological solutions is not necessarily guaranteed
during the whole cosmological history. 
Indeed, for the cubic-order generalized Galileon and its extensions, 
the Laplacian instabilities arise during the transition from the 
bouncing/genesis period to the subsequent 
stage \cite{Cai12,Koehn,Battarra,Qiu15,Libanov}. 
In Ref.~\cite{Koba16}, it was shown that this conclusion 
also holds for full Horndeski theories. 
In what follows, we first revisit the no-go argument 
in Horndeski theories for the absence of stable nonsingular 
solutions throughout the cosmic history
and then discuss what happens 
in GP and SVT theories. 

\subsubsection{Horndeski theories}

In Horndeski theories with the matter perfect fluid, there are two 
dynamical scalar DOFs. In the unitary gauge (\ref{gauge3}), 
these DOFs are characterized by the perturbations 
\be
{\cal X}^t=\left( {\cal R}_{\phi}, 
\delta \rho_{\rm u}/k \right)\,.
\ee
After integrating out nondynamical DOFs, 
the second-order action of scalar perturbations 
is of the form (\ref{Ss2}) with $2 \times 2$ matrices 
${\bm K}$, ${\bm G}$, ${\bm M}$, ${\bm B}$. 
In the small-scale limit, the stability conditions of 
$\delta \rho_{\rm u}$ are the same as those 
in SVT theories, i.e., $\rho_m+P_m>0$ and $c_m^2>0$. 
For the perturbation ${\cal R}_{\phi}$, the no-ghost 
condition corresponds to
\be
q_s^{{\rm (u)}} \equiv 
q_t \left( 3+\frac{4q_t w_4}{w_1^2} \right)>0\,,
\ee
where $q_s^{{\rm (u)}}$ is equivalent to the matrix 
component $K_{22}$ in Eq.~(\ref{Kcom}) with $w_2=w_5=w_8=0$. 
In Horndeski theories, the product of $q_s^{{\rm (u)}}$ and 
the scalar propagation speed squared $c_{s}^2$ is equivalent 
to $G_{22}$ in Eq.~(\ref{Gcom}) without the term $2E_3^2/q_v$. 
Then, we obtain the following relation 
\be
\frac{1}{a} \frac{d}{dt} \left( aE_2 \right)=
q_s^{{\rm (u)}} c_s^2+q_t c_t^2
+\frac{2q_t^2 (\rho_m+P_m)}{w_1^2}\,,
\label{qscs}
\ee
where 
\be
E_2=-\frac{2q_t^2}{w_1}\,.
\ee

In the bouncing Universe, the scale factor $a(t)$ reaches a positive 
minimum at the bounce and 
it approaches a positive constant or diverges 
in the asymptotic past ($t \to -\infty$) and future 
($t \to \infty$). 
The genesis model corresponds to the case in which 
the scale factor and its time derivative are finite 
for all $-\infty<t<\infty$.  
Since we require that the perturbations are prone 
to neither ghost nor Laplacian instabilities, 
the three terms on the right hand side of Eq.~(\ref{qscs}) 
are positive. Then, the following inequality holds
\be
\frac{1}{a} \frac{d \xi}{dt} 
>q_t c_t^2>0\,,
\label{E2re}
\ee
where 
\be
\xi \equiv aE_2=-\frac{2aq_t^2}{w_1}\,.
\label{xidef}
\ee
Integrating Eq.~(\ref{E2re}) from $t=t_i$ to $t=t_f~(>t_i)$, 
we obtain
\be
\xi_f-\xi_i
>\int_{t_i}^{t_f} a\,q_t c_t^2\,dt>0\,.
\label{aE2}
\ee

In the following, we consider the case in which the quantity $q_t c_t^2$ 
does not approach 0 for $t_i \to -\infty$ and $t_f \to \infty$.
The limit $q_t c_t^2 \to 0$ corresponds to either $q_t \to 0$ 
or $c_t^2 \to 0$. For $q_t \to 0$, the strong coupling problem 
of tensor perturbations arises. In another limit $c_t^2 \to 0$ 
the gradient term in Eq.~(\ref{St2}) vanishes, so nonlinear 
contributions to the tensor action are out of control.
{}From the view point of quantum field theory, the leading-order 
solution to $h_i$ corresponding to the Bunch-Davies vacuum is proportional to 
$1/(c_t k)^{3/2}$ \cite{GWDT}, which diverges for $c_t \to 0$.

Since $q_t c_t^2$ does not decrease toward 0 in the asymptotic future, 
the integral in Eq.~(\ref{aE2}) is a positive growing 
function of $t_f$. Hence the consistency of 
Eq.~(\ref{aE2}) demands that $\xi_f>0$ 
for sufficiently large $t_f$. 
The integral also increases toward the asymptotic 
past ($t_i \to -\infty$), so we require the condition $\xi_i<0$.
Then, the function $\xi$ crosses 0 at some time between 
$-\infty<t<\infty$, which correspond to $a=0$ from Eq.~(\ref{xidef}). 
This behavior is at odds with the nonsingular bouncing/genesis 
cosmology in which $a>0$ throughout the cosmological evolution. 

The no-go argument given above has been proven in the unitary gauge. 
The same argument also holds for other gauges in which 
the perturbations are well defined at the bounce.
By choosing ${\cal R}_{\phi}=\zeta-H \delta \phi_{\rm N}/\dot{\phi}$ 
and $\delta \rho_{\rm N}/k$ as dynamical perturbations in 
the Newtonian gauge, the quantities $q_s$ and $c_s^2$ 
are equivalent to those in the unitary gauge (see the discussion at the 
end of Sec.~\ref{Newsec}). In this case the same relation as
Eq.~(\ref{qscs}) holds, so the no-go argument given above is also applied 
to the Newtonian gauge for the dynamical perturbations 
${\cal R}_{\phi}$ and $\delta \rho_{\rm N}$.
Hence, the absence of consistent bouncing solutions in Horndeski theories 
is not an artifact of the gauge choice,
but it is a real unavoidable physical problem. 
In the following, we will see how this problem can be naturally 
avoided in GP and SVT theories due to
the presence of intrinsic vector modes.

\subsubsection{GP theories}

The crucial point of the no-go argument in Horndeski 
theories is that, besides the term $q_t c_t^2$, all 
the other terms on the right hand side of Eq.~(\ref{qscs}) 
are positive for the absence of ghost and Laplacian 
instabilities. Let us consider GP theories in the presence 
of a matter perfect fluid.  
Choosing the uniform vector gauge (\ref{gaugev}), 
the dynamical scalar DOFs are given by 
\be
{\cal X}^t=\left( {\cal R}_{\psi}, 
\delta \rho_{\rm v}/k \right)\,.
\ee

Since the scalar-field perturbation $\delta \phi$ is absent in 
GP theories, the computation of 
$q_s^{(\rm v)}$ and $c_s^2$ in the uniform vector gauge 
($\psi=0$) is analogous to that of $q_s^{(\rm u)}$ and $c_s^2$
in Horndeski theories with the choice of unitary gauge ($\delta \phi=0$).
For the matter perturbation $\delta \rho_{\rm v}$, the conditions 
for the absence of ghost and Laplacian instabilities are given by 
$\rho_m+P_m>0$ and $c_m^2>0$. 
In GP theories, there are the following relations 
\be
w_1=w_2-2Hq_t\,,\qquad 
w_4=w_5+\frac{3}{2}H \left( w_1+w_2 \right)\,,\qquad 
w_8=3Hw_1-2w_4\,.
\label{w148}
\ee
For the perturbation ${\cal R}_{\psi}$, the ghost is absent for 
\be
q_s^{(\rm v)} \equiv \frac{q_t (3w_2^2+4w_5 q_t)}
{(2H q_t+w_2)^2}>0\,,
\label{qsv}
\ee
which is equivalent to $K_{22}$ in Eq.~(\ref{Kcom}) 
after the substitution of Eq.~(\ref{w148}). 
Since the Hubble parameter $H$ does not appear in 
the numerator of $q_s^{(\rm v)}$, the strong coupling 
problem does not arise at $H=0$.
The product of $q_s^{(\rm v)}$ and $c_s^2$ reduces to 
the same form as $G_{22}$ in Eq.~(\ref{Gcom}) with the 
particular relations (\ref{w148}). 
Then, it follows that 
\be
\frac{1}{a} \frac{d}{dt} \left( aE_2 \right)=
q_s^{{\rm (v)}} c_s^2+q_t c_t^2
+\frac{2q_t^2 (\rho_m+P_m)}{(2Hq_t+w_2)^2}
-\frac{2E_3^2}{q_v}\,,
\label{qscsGP}
\ee
where 
\be
E_2=\frac{2q_t^2}{2Hq_t+w_2}\,,\qquad 
E_3=\frac{1}{4H} \left[ \frac{w_2 (2H q_t-w_2)}
{A_0(2Hq_t+w_2)}-w_6 \right]\,.
\ee
Compared to the relation (\ref{qscs}) in Horndeski theories, 
there is the additional term $-2E_3^2/q_v$ in Eq.~(\ref{qscsGP}). 
This new term arises from the existence of intrinsic vector modes. 
Since $q_v>0$ for the absence of vector ghosts, 
the term $-2E_3^2/q_v$ needs to be negative. 
Then, unlike Horndeski theories, 
the right hand side of Eq.~(\ref{qscsGP}) is no longer
bounded from below with the minimum value $q_tc_t^2$. 

Integration of Eq.~(\ref{qscsGP}) from $t=t_i$ to $t=t_f$ leads to 
\be
\xi_f-\xi_i=\int_{t_i}^{t_f} a \left[ 
q_s^{{\rm (v)}} c_s^2+q_t c_t^2
+\frac{2q_t^2 (\rho_m+P_m)}{(2Hq_t+w_2)^2}
-\frac{2E_3^2}{q_v} \right] dt\,,
\label{xif}
\ee
where $\xi=aE_2=2aq_t^2/(2Hq_t+w_2)$.
If the contribution $-2E_3^2/q_v$ to the square bracket 
of Eq.~(\ref{xif}) dominates over the other terms
in the asymptotic past ($t_i \to -\infty$), then the integral goes to $-\infty$ and 
hence $\xi_i>0$. If the term $-2E_3^2/q_v$ is subdominant to 
$q_tc_t^2~(>0)$ in the asymptotic future ($t_f \to \infty$), 
the integral grows toward $\infty$ and hence $\xi_f>0$.
In this case, it is possible to have $\xi>0$ throughout 
the cosmological evolution.
This means that, in GP theories, there is a possibility 
for realizing nonsingular bouncing/genesis solutions where
the scale factor is always in the region $a>0$. 
This is a very promising property of GP theories for bouncing 
solutions compared to Horndeski theories.

\subsubsection{SVT theories}

In SVT theories, there are two scalar propagation speed squares 
given by Eq.~(\ref{cs}) and hence
\ba
& &
q_s c_{s1}^2+q_s c_{s2}^2={\cal F}_s\,,\\
& &
q_s c_{s1}^2 c_{s2}^2={\cal G}_s\,.
\label{qsSVT}
\ea
The positivities of ${\cal F}_s$ and ${\cal G}_s$ are required 
to avoid ghost and Laplacian instabilities of scalar perturbations.
{}From Eq.~(\ref{qsSVT}), it follows that 
\be
G_{11}G_{22}=q_s c_{s1}^2 c_{s2}^2+G_{12}^2>0\,,
\ee
which means that either 
(i) $G_{11}>0$ and $G_{22}>0$, or 
(ii) $G_{11}<0$ and $G_{22}<0$. 
In the unitary gauge, the expressions of $G_{11}$ and $G_{22}$ 
have been derived in Eq.~(\ref{Gcom}), so that 
\ba
& &
\frac{1}{a} \frac{d}{dt} \left( aE_1 \right)
=G_{11}-\frac{\alpha_2}{2}-\frac{2E_1^2}{q_v}
+\frac{w_2^2 (\rho_m+P_m)}{2A_0^2 (w_1-2w_2)^2}\,,
\label{aE1eq} \\
& &
\frac{1}{a} \frac{d}{dt} \left( aE_2 \right)
=G_{22}+q_t c_t^2-\frac{2E_3^2}{q_v}
+\frac{2q_t^2 (\rho_m+P_m)}{(w_1-2w_2)^2}\,,
\label{aE2eq}
\ea
where $E_1, E_2, E_3$ are defined by Eq.~(\ref{E123}).
The tachyonic instability of vector perturbations can be avoided 
for $\alpha_2>0$, but this condition is not obligatory compared to 
conditions for the absence of ghost and Laplacian instabilities. 
For $G_{11}>0$ and $G_{22}>0$, the situation is analogous to 
what we discussed in GP theories. 
The intrinsic vector-mode contributions $-2E_1^2/q_v$ and 
$-2E_3^2/q_v$ to Eqs.~(\ref{aE1eq}) and (\ref{aE2eq}), which 
are required to be negative, allow the possibility for evading 
the no-go argument in Horndeski theories, in such a way that the 
quantities $aE_1$ and $aE_2$ can remain positive throughout 
the cosmological evolution. 
When $G_{11}<0$ and $G_{22}<0$, 
the no-go statement does not hold either.
Thus, in SVT theories, it would be possible to realize 
nonsingular bouncing/genesis solutions without theoretical 
pathologies. We note that such nonsingular solutions should be
constructed to satisfy the conditions ${\cal F}_s>0$ and 
${\cal G}_s>0$ besides $q_s>0$, without having the 
behavior $q_t c_t^2 \to 0$ in the asymptotic past and future.
 
\section{Application to dark energy}
\label{conmodelsec}

In this section, we apply the gauge-ready formulation 
of Sec.~\ref{scasec} to the case in which the scalar field 
$\phi$ and the vector field $A_{\mu}$ are the source 
for the late-time cosmic acceleration. 
For the matter action ${\cal S}_m$, we consider a nonrelativistic 
perfect fluid satisfying $P_m \simeq 0$ and $c_m^2 \simeq 0$. 
We are interested in observables relevant to the evolution 
of matter perturbations and gravitational potentials to test 
dark energy models in SVT theories with the measurements 
of redshift-space distortions, weak lensing, and CMB. 

{}From Eqs.~(\ref{eqdrho}) and (\ref{veq}), 
the matter perturbation $\delta \rho_m$ and the velocity 
potential $v$ obey
\ba
& &
\dot{\delta \rho}_m+3H \delta \rho_m
+\rho_m \left[ 3 \dot{\zeta}+\frac{k^2}{a^2}
\left( v+\chi-a^2\dot{E} \right) \right]=0\,,
\label{delrhoeq} \\
& &
\dot{v}=\alpha\,.
\label{veqa}
\ea
Taking the time derivative of Eq.~(\ref{delrhoeq}) and using 
Eq.~(\ref{veqa}), the gauge-invariant density contrast 
$\delta_m=\delta \rho_m/\rho_m+3Hv$ satisfies 
\be
\ddot{\delta}_m+2H \dot{\delta}_m+\frac{k^2}{a^2}\Psi
=3 \left( \ddot{\cal B}+2H\dot{\cal B} \right)\,,
\ee
where ${\cal B}=Hv-\zeta$, and $\Psi$ is the gauge-invariant 
gravitational potential defined in Eq.~(\ref{PsiPhi}). 
We relate the Newtonian gravitational potential $\Psi$ and 
the weak lensing potential $\psi_{\rm eff}=\Phi-\Psi$ with $\delta_m$, as 
\be
\frac{k^2}{a^2} \Psi=-4\pi G \mu \rho_m \delta_m\,,\qquad 
\frac{k^2}{a^2} \psi_{\rm eff}
=8\pi G \Sigma \rho_m \delta_m\,,
\label{muSig}
\ee
where $\mu$ and $\Sigma$ are dimensionless quantities, 
and $G$ is the Newton gravitational constant.
The quantity $\Sigma$ can be expressed as 
\be
\Sigma=\frac{1+\eta}{2}\mu\,,\qquad \eta \equiv 
-\frac{\Phi}{\Psi}\,,
\label{eta}
\ee
where $\eta$ is dubbed the gravitational slip parameter.
The deviations of $\mu$ and $\Sigma$ from 1 lead to  
the modified evolution of $\Psi, \psi_{\rm eff}$, 
and $\delta_m$ compared to the case of GR. 

In Ref.~\cite{Kasedark}, the calculations of $\mu$ and $\Sigma$ 
were performed by choosing the flat gauge 
(\ref{flat}), but the separation of those quantities 
between tensor, vector, and scalar contributions 
is not transparent. Since $\zeta=0$ in the flat gauge, the quantities 
$q_t$ and $c_t^2$ do not explicitly appear as coefficients 
of the flat-gauge Lagrangians (\ref{L1})-(\ref{L3}). 
As we see in Eqs.~(\ref{Lze})-(\ref{LE}), this situation is 
different in other gauges where $\zeta$ does not vanish. 
In the following, we choose the unitary gauge given by 
\be
\delta \phi=0\,,\qquad E=0\,.
\ee
We employ the quasi-static approximation on sub-horizon 
scales \cite{quasi1,quasi2,DKT}, 
under which the dominant contributions to the perturbation equations 
of motion are those containing $k^2/a^2$ and $\delta \rho_m$. 
In doing so, we introduce the dimensionless quantities:
\ba
& &
\epsilon_H \equiv \frac{\dot{H}}{H^2}\,,\qquad \Omega_m \equiv 
\frac{\rho_m}{3H^2 q_t}\,,\qquad \epsilon_A \equiv 
\frac{\dot{A}_0}{HA_0}\,,\qquad
\alpha_{\rm B} \equiv -\frac{w_1-2w_2+2H q_t}
{2H q_t}\,,\qquad
x_2 \equiv \frac{w_2}{H q_t}\,,\qquad
x_6 \equiv \frac{A_0 w_6}{H q_t}\,,\nonumber \\
& &
y_{\rm B} \equiv \frac{\dot{\alpha}_{\rm B}}
{H \alpha_{\rm B}}\,,\qquad 
y_2 \equiv \frac{\dot{x}_2}{H x_2}\,,\qquad 
y_6 \equiv  \frac{\dot{x}_6}{H x_6}\,,\qquad 
\varphi_{\rm u} \equiv \frac{H}{A_0}\psi_{\rm u}\,,
\qquad
\beta_2 \equiv \frac{A_0^2 \alpha_2}{H^2 q_t}\,,
\qquad 
q_r \equiv \frac{q_t}{A_0^2 q_v}\,,
\label{EFTdef}
\ea
and $\alpha_{\rm M}$ defined by Eq.~(\ref{alM}). 
If we switch off the vector field, the parameters $\alpha_{\rm M}$ 
and $\alpha_{\rm B}$ reduce to those introduced in 
Horndeski theories in Ref.~\cite{Bellini}, which 
represent the running of gravitational constant and 
the kinetic mixing between the scalar field and gravity, 
respectively \cite{Bellini}. 
In Appendix~\ref{aKaT}, we also show the correspondence with 
other dimensionless parameters introduced in Ref.~\cite{Bellini} 
(such as $\alpha_{\rm T}$ and $\alpha_{\rm K}$).

In the unitary gauge, there are three dynamical perturbations  
$\psi_{\rm u}=\psi, {\cal R}_{\phi}=\zeta$, 
and $\delta \rho_{\rm u}=\delta \rho_{m}$ 
with the gravitational potentials   
$\Psi=\alpha+\dot{\chi}$ and $\Phi=\zeta+H \chi$.
Applying the quasi-static approximation to 
Eqs.~(\ref{eqdA}) and (\ref{eqalpha}), respectively, 
it follows that 
\ba
{\cal Y}
&=& \frac{A_0 w_6-w_2}{A_0}\psi_{\rm u}-2w_2 \chi
-4A_0 \alpha_3 {\cal R}_{\phi} \label{Ycond} \\
&=& 2\left( q_t-2A_0 \alpha_3 \right){\cal R}_{\phi}
+w_6 \psi_{\rm u}-w_1 \chi-\frac{a^2}{k^2} \delta \rho_{\rm u}\,.
\label{Ycon}
\ea
Then, the term ${\cal Y}$ can be eliminated to give
\ba
\delta \rho_{\rm u}
&=& -\frac{k^2}{a^2} \left[ (w_1-2w_2)\chi-2q_t {\cal R}_{\phi}
-\frac{w_2}{A_0} \psi_{\rm u} \right] \label{quasi1d} \\
&=&
\frac{k^2}{a^2}q_t \left[2 \left( 1+\alpha_{\rm B} \right) \Phi 
-2\alpha_{\rm B}{\cal R}_{\phi}+x_2  \varphi_{\rm u} \right]\,.
\label{quasi1}
\ea
We take the time derivative of Eq.~(\ref{quasi1d}) and substitute 
$\dot{\delta \rho}_{\rm u}$ and $\delta \rho_{\rm u}$ into Eq.~(\ref{eqdrho}). 
In doing so, we exploit Eq.~(\ref{eqchi}) to remove the perturbation $v$ from 
Eq.~(\ref{eqdrho}) and eliminate the time derivative $\dot{\psi}_{\rm u}$ 
in $\dot{\delta \rho}_{\rm u}$ by using Eqs.~(\ref{Ydef}) and (\ref{Ycond}). 
This process finally leads to the disappearance of $\dot{\cal R}_{\phi}$. 
After replacing the combination $\alpha+\dot{\chi}$ with $\Psi$, 
we obtain
\be
b_1 \Phi+4 \left( 1+\alpha_{\rm B} \right) \Psi
+b_2 {\cal R}_{\phi}+b_3  \varphi_{\rm u}=0\,,
\label{quasi2}
\ee
where 
\ba
b_1 &=&4 \left( 1+\alpha_{\rm B} \right) \left( 1+\alpha_{\rm M}+\epsilon_H \right)
+4 \alpha_{\rm B}y_{\rm B}+6\Omega_m-2x_2^2q_r\,,\\
b_2 &=& 4 \left( 1+\alpha_{\rm M} \right)
-x_2 \left( x_2+x_6 \right)q_r-b_1\,,\\
b_3 &=&x_2 \left[ 2(1+\alpha_{\rm M}+\epsilon_H-\epsilon_A+y_2)
-\left( x_2-x_6 \right)q_r \right]\,.
\ea
We also differentiate Eq.~(\ref{Ycond}) with respect to $t$ and eliminate 
the terms $\dot{\cal Y}$ and ${\cal Y}$ from Eq.~(\ref{calYeq}).
This gives
\be
-2b_3 \Phi-4x_2 \Psi
+b_4 {\cal R}_{\phi}+b_5  \varphi_{\rm u}=0\,,
\label{quasi3}
\ee
where 
\ba
b_4 &=&2 x_2 \left( 1+\alpha_{\rm M}
+2\epsilon_H-\epsilon_A
+y_2 \right)
-2x_6 \left( 1+\alpha_{\rm M}-\epsilon_A+y_6 
\right)-\left( x_2-x_6 \right)^2q_r\,,\\
b_5 &=& -2 x_2 \left( 1+\alpha_{\rm M}
+\epsilon_H-2\epsilon_A
+y_2 \right)
+2x_6 \left( 1+\alpha_{\rm M}
+\epsilon_H-2\epsilon_A+y_6 
\right)+ \left( x_2-x_6 \right)^2q_r+4\beta_2\,.
\ea
Substituting Eq.~(\ref{Ycond}) into Eq.~(\ref{Bianchi}), 
it follows that 
\be
2 \left( b_1+b_2 \right) \Phi+8 \Psi+b_6 {\cal R}_{\phi}
+\left( 2b_3-b_4 \right) \varphi_{\rm u}=0\,,
\label{quasi4}
\ee
where 
\be
b_6=8 \left( c_t^2-1-\alpha_{\rm M} \right)
+\left( x_2^2-x_6^2 \right) q_r\,.
\ee

Solving Eqs.~(\ref{quasi1}), (\ref{quasi2}), (\ref{quasi3}), 
and (\ref{quasi4}) for $\Phi, \Psi, {\cal R}_{\phi}$, and 
$\varphi_{\rm u}$, we obtain 
\ba
\Phi &=& 
\frac{4[b_2(2x_2 b_3-x_2 b_4+2b_5)+b_3(2 \alpha_{\rm B}b_4-x_2 b_6)
-(1+\alpha_{\rm B})(b_4^2+b_5b_6)]}
{\Delta} \frac{a^2}{k^2} \delta \rho_{\rm u}\,,\\
\Psi &=& 
-\frac{b_1(2b_2b_5-b_4^2-b_5b_6)+2b_2(b_2b_5+2b_3^2-2b_3b_4)-2b_3^2 b_6}
{\Delta} \frac{a^2}{k^2} \delta \rho_{\rm u}\,,\\
{\cal R}_{\phi} &=&
\frac{4[x_2(b_1b_4+2b_2b_3)+2\alpha_{\rm B}
(b_1b_5+b_2b_5+2b_3^2-b_3b_4)+2b_2b_5-2b_3b_4]}
{\Delta} \frac{a^2}{k^2} \delta \rho_{\rm u}\,,\\
\varphi_{\rm u} &=&
-\frac{4[x_2(2b_1b_2-b_1b_6+2b_2^2)+2\alpha_{\rm B}b_1b_4
+2(1+\alpha_{\rm B})(b_2b_4+b_3b_6)-4b_2b_3]}
{\Delta} \frac{a^2}{k^2} \delta \rho_{\rm u}\,,
\ea
where the determinant $\Delta$ can be expressed in terms of 
the quantity $q_s^{{\rm (u)}} c_{s1}^2 c_{s2}^2$, as 
\be
\Delta=\frac{512A_0^2(1+\alpha_{\rm B})^2}
{H^2q_t}q_s^{{\rm (u)}} c_{s1}^2 c_{s2}^2\,.
\label{Delqs}
\ee
For the derivation of the relation (\ref{Delqs}), we used the fact that 
$q_s^{{\rm (u)}} c_{s1}^2 c_{s2}^2=G_{11}G_{22}-G_{12}^2$ with 
$G_{11}, G_{22}, G_{12}$ given by Eq.~(\ref{Gcom}).
Since the approximation $\delta_m \simeq \delta \rho_{\rm u}/\rho_m$ 
holds for the perturbations deep inside the Hubble radius, 
the quantities $\mu$ and $\eta$ defined in Eqs.~(\ref{muSig}) 
and (\ref{eta}) yield
\ba
\mu &=& 
\frac{H^2 q_t[b_1(2b_2b_5-b_4^2-b_5b_6)+2b_2(b_2b_5+2b_3^2-2b_3b_4)-2b_3^2 b_6]}
{2048 \pi G A_0^2 (1+\alpha_{\rm B})^2q_s^{{\rm (u)}} c_{s1}^2 c_{s2}^2}\,,\label{muf}\\
\eta &=&
\frac{4[b_2(2x_2 b_3-x_2 b_4+2b_5)+b_3(2 \alpha_{\rm B}b_4-x_2 b_6)
-(1+\alpha_{\rm B})(b_4^2+b_5b_6)]}{b_1(2b_2b_5-b_4^2-b_5b_6)+2b_2(b_2b_5+2b_3^2-2b_3b_4)-2b_3^2 b_6}\,.\label{etaf}
\ea
We note that $\mu$ contains the matter density parameter 
$\Omega_m$ through $b_1$ and $b_2$. 
{}From Eq.~(\ref{Gcom}), the product 
$q_s^{{\rm (u)}} c_{s1}^2 c_{s2}^2=G_{11}G_{22}-G_{12}^2$ also 
contains the term linear in $\Omega_m$.
After using this relation to eliminate $\Omega_m$ from 
Eq.~(\ref{muf}), we find that $\mu$ is expressed in the form 
\be
\mu=\mu_0 \left[ 1+\frac{\mu_1^2}{\mu_2\,q_s^{{\rm (u)}} c_{s1}^2 c_{s2}^2} 
\right]\,,
\label{muex}
\ee
where 
\ba
&&
\mu_0=\frac{[2(b_1+b_2)+b_6]\xi_0}{8\pi Gq_t}
\left[8\xi_0-\frac{q_vA_0^2}{2}\left\{\left(b_1+b_2+\frac{b_6}{2}\right)x_2
-2\left(b_3-\frac{b_4}{2}\right)\right\}^2\right]^{-1}\,,
\label{mu0}\\
&&
\mu_1=\frac{Hq_t q_v
\left[x_2\{(b_1+b_2)b_4+b_3b_6\}
+(2b_3-b_4)\{2\aB b_3-(1+\aB)b_4\}
+b_5\{2\aB(b_1+b_2)+(1+\aB)b_6\}\right]}{32(1+\aB)}\,,
\label{mu1}\\
&&
\mu_2=q_v\xi_0\,, 
\label{mu2}
\ea
with 
\be
\xi_0\equiv\frac{A_0^2 q_v}{8}\left[\left(b_1+b_2+\frac{b_6}{2}\right)b_5
+2\left(b_3-\frac{b_4}{2}\right)^2\right]\,. 
\label{defxi0}
\ee
In Eqs.~(\ref{mu0})-(\ref{defxi0}), the quantities $b_1$ and $b_2$ appear only 
through the combination $b_1+b_2$, which does not contain $\Omega_m$.

In GR we have $\mu_0=1$ and $\mu_1=0$, but in SVT theories the 
modifications arising from tensor, vector, scalar sectors generally 
lead to $\mu_0 \neq 1$ and $\mu_1 \neq 0$. 
Since $b_6$ contains $c_t^2$, the term $\mu_0$ depends on 
$q_t, c_t^2, q_v$, i.e., the quantities associated with the stabilities 
of tensor and vector perturbations.
The second term in the square bracket of Eq.~(\ref{muex}) 
is dependent on $q_t, c_t^2, q_v, q_s, c_{s1}^2, c_{s2}^2$, 
so that this characterizes the matter interaction 
with tensor, vector, and scalar sectors. 
Thus, the separation of $\mu$ between 
tensor, vector, and scalar contributions is clear in the unitary gauge, 
but this is not the case for the
flat gauge chosen in Ref.~\cite{Kasedark}. 
Even though $\mu$ and $\eta$ are gauge-invariant quantities, the unitary gauge 
is more convenient than the flat gauge for this problem in that 
the physical interpretation of gravitational interactions becomes transparent.
Provided the ghost and Laplacian instabilities are absent in the scalar sector, 
the quantity $\mu_1^2/(q_s^{{\rm (u)}} c_{s1}^2 c_{s2}^2)$ is positive. 
Depending on the sign of $\mu_2$, we have either (a) $\mu>\mu_0$ 
for $\mu_2>0$, or (b) $\mu<\mu_0$ for $\mu_2<0$.

{}From the GW170817 event \cite{Abbott} together with the gamma-ray burst 
GRB 170817A \cite{GRB}, 
the speed of tensor perturbations needs to be very close to 1 
for the redshift $z<0.009$. In the following, we focus on  
SVT theories satisfying the condition 
\be
c_t^2=1\,.
\label{ctcon}
\ee
If we do not admit any tuning among functions in 
Eq.~(\ref{ct}), the couplings are constrained to be 
\be
G_4=G_4(\phi)\,,\qquad G_5=0\,,\qquad 
f_4=0\,,\qquad f_5=0\,,
\label{conmo1}
\ee
with all the other functions like $f_6(\phi, X_1)$ allowed.
Note that the $\phi$ dependence in $f_4$ has been absorbed into $G_4(\phi)$.
For the couplings (\ref{conmo1}) the quantity $\alpha_3$ 
defined by Eq.~(\ref{al3}) vanishes, 
so there is the particular relation 
$w_2=-A_0 w_6$ from Eq.~(\ref{property1}). 
Then, the following relations hold
\be
x_2=-x_6\,,\qquad y_2=y_6\,,
\ee
under which we have 
\be
b_1+b_2=4 \left( 1+\alpha_{\rm M} \right)\,,\qquad
b_1+b_2+\frac{b_6}{2}=4\,, \qquad 
b_3-\frac{b_4}{2}=0\,.
\ee
Substituting these relations 
into Eqs.~(\ref{mu0}), (\ref{mu1}) and (\ref{mu2}), 
we obtain 
\be
\mu_0=\frac{b_5}{8\pi G q_t (b_5-2x_2^2)}\,,
\qquad 
\mu_1=\frac{Hq_t q_v[b_3 x_2+b_5(\alpha_{\rm B}
-\alpha_{\rm M})]}
{4(1+\alpha_{\rm B})}\,,\qquad 
\mu_2=\frac{A_0^2 q_v^2 b_5}{2}\,,
\label{mu012}
\ee
with $\xi_0=A_0^2 q_v b_5/2$, and 
\be
b_5=4\beta_2+4x_2^2 q_r
-4x_2 \left( 1+\alpha_{\rm M}
+\epsilon_H-2\epsilon_A
+y_2 \right)\,.
\label{b5}
\ee
When $b_5>0$, the positivity of $\mu_0$
requires that 
\be
b_5>2x_2^2\,.
\ee
In this case we have $\mu>\mu_0>1/(8\pi G q_t)$, 
so the gravitational interaction is stronger than that 
in GR for linear cosmological perturbations.

If $b_5<0$, it follows that $\mu<\mu_0<1/(8\pi G q_t)$. 
Then, the gravitational interaction is weaker than that in GR. 
If the vector mass squared is positive ($\beta_2>0$), 
the first two terms on the right hand side of Eq.~(\ref{b5}) 
are positive under the absence of tensor and vector ghosts.
Then, the only possibility for realizing $b_5<0$ is that 
the contribution $-4x_2(1+\alpha_{\rm M}+\epsilon_H-2\epsilon_A+y_2)$ in Eq.~(\ref{b5}) 
is negative and it overwhelms other 
positive terms. It remains to be seen whether 
this behavior is possible for concrete dark energy models 
in the framework of SVT theories.

Finally, we further specify cubic couplings in the form 
\be
f_3=f_3(\phi)\,,\qquad \tilde{f}_3=0\,,
\label{f3phi}
\ee
in addition to the functions (\ref{conmo1}). 
Since $w_2=0$ in this case, we have $x_2=0$ and 
$b_5=4\beta_2=4A_0^2 \alpha_2/(H^2 q_t)$.
Substituting these relations into Eq.~(\ref{mu012}), 
the quantity (\ref{muex}) reduces to  
\be
\mu = \frac{1}{8\pi G q_t} \left[ 1+\alpha_2
\frac{q_t(\alpha_{\rm B}-\alpha_{\rm M})^2}
{2(1+\alpha_{\rm B})^2 q_s^{{\rm (u)}} c_{s1}^2 c_{s2}^2}
\right]\,.\label{muf2}
\ee
{}From Eq.~(\ref{etaf}),  the gravitational slip 
parameter yields
\be
\eta= \frac{2(1+\alpha_{\rm B})^2q_s^{{\rm (u)}} c_{s1}^2 c_{s2}^2
+\alpha_2 q_t \alpha_{\rm B} (\alpha_{\rm B}-\alpha_{\rm M})}
{2(1+\alpha_{\rm B})^2q_s^{{\rm (u)}} c_{s1}^2 c_{s2}^2
+\alpha_2 q_t (\alpha_{\rm B}-\alpha_{\rm M})^2}\,.
\label{etaf2}
\ee
Now, the explicit dependence on $q_v$ disappears from 
$\mu$ and $\eta$. 
The intrinsic vector modes implicitly affect $\mu$ and $\eta$
through the dependence of  $c_{s1}^2$ and $c_{s2}^2$ on $q_v$. 
Provided that $\alpha_{\rm B}$ and $\alpha_{\rm M}$ do not vanish 
with $\alpha_{\rm B} \neq \alpha_{\rm M}$, $\mu$ differs from the 
value $1/(8\pi G q_t)$. This property is analogous to what happens 
in Horndeski theories with $c_t^2=1$, 
in which case the braiding parameter $\alpha_{\rm B}$ 
and the running parameter $\alpha_{\rm M} $ of $q_t$ lead to the  
gravitational interaction different from that in GR 
(see, e.g., Eqs.~(3.36) and (3.37) of 
Ref.~\cite{Kase18}).

Compared to the values of $\mu$ and $\eta$ in 
Horndeski theories, the vector mass squared $\alpha_2$ 
appears in Eqs.~(\ref{muf2})-(\ref{etaf2}), in addition to the 
presence of the product $c_{s1}^2 c_{s2}^2$ instead of a single 
sound speed squared $c_s^2$. If the condition 
\be
\alpha_2=f_{2,X_3}>0
\label{al2con}
\ee
is satisfied, the gravitational interaction is enhanced 
(i.e., $\mu>1/(8\pi G q_t)$) compared to that in GR under the stability 
conditions $q_t>0$ and $q_s^{{\rm (u)}} c_{s1}^2 c_{s2}^2>0$.
Since $w_2=0$ in the present theory, the matrix component $K_{11}$ 
in the unitary gauge reduces to $w_5/A_0^2$. 
On using the background Eq.~(\ref{back4}), i.e., 
$(f_{2,X_2}+4f_{3,\phi})\dot{\phi}=-2f_{2,X_3}A_0$ to simplify $w_5$, it follows that 
\be
K_{11}=\frac{1}{8} \left( 4f_{2,X_3}
+f_{2,{X_2 X_2}}\dot{\phi}^2+4f_{2,{X_2 X_3}} \dot{\phi}A_0
+4f_{2,{X_3 X_3}}A_0^2
\right)\,.
\label{K11ex}
\ee
For the theories in which $f_2$ contains only linear functions 
of $X_2$ and $X_3$, we have $K_{11}=f_{2,X_3}/2=\alpha_2/2$ and 
hence $K_{11}$ and $\alpha_2$ have the same sign.
For the tachyonic vector mass squared $\alpha_2<0$, the negative 
value of $K_{11}$ implies that $K_{22}$ needs to be negative to 
satisfy the condition $q_s=K_{11}K_{22}-K_{12}^2>0$. 
In this case the scalar ghost appears, so we require the condition 
$\alpha_2>0$. 
Then the gravitational interaction is stronger than that in GR. 
The only possibility for realizing $\mu<1/(8\pi G q_t)$ is to introduce 
nonlinear terms in $X_2$ and $X_3$ which overwhelm the negative 
term $f_{2,X_3}$ in $K_{11}$. 
Since the last three terms in the bracket of Eq.~(\ref{K11ex}) 
contain the time-dependent fields $\dot{\phi}$ and $A_0$, 
we generally require the tuning of functions 
to keep the condition $K_{11}>0$ throughout the cosmological evolution for $f_{2,X_3}<0$.

The gravitational slip parameter (\ref{etaf2}) is generally different from 1, but there 
are specific theories in which $\eta$ is equivalent to 1. 
They are characterized by three cases: (i) $\alpha_2=0$, 
(ii) $\alpha_{\rm B}=\alpha_{\rm M}$, and (iii) $\alpha_{\rm M}=0$.
In cases (i) and (ii) the quantity (\ref{muf2}) simply reduces to 
$\mu=1/(8\pi G q_t)$, but in case (iii) the second term in the square bracket 
of Eq.~(\ref{muf2}) 
does not vanish for $\alpha_{\rm B} \neq 0$. 
For example, the cubic coupling $G_3(X_1)$ gives rise to a 
nonvanishing contribution to $\alpha_{\rm B}$. 
Apart from the specific cases (i), (ii), (iii), the quantity 
$\Sigma=(1+\eta)\mu/2$ differs from $\mu$.  
We note that the quartic nonminimal coupling $G_4(\phi)$ 
affects $\mu$ and $\Sigma$ through 
the nonvanishing contributions to $\alpha_{\rm M}$ 
as well as to $\alpha_{\rm B}$.

\section{Conclusions}
\label{concludesec}

In parity-invariant SVT theories with broken $U(1)$ gauge symmetry, 
we developed the gauge-ready formulation of scalar 
cosmological perturbations by taking into account a matter perfect fluid. 
In such theories, there are three scalar DOFs arising from a scalar 
field $\phi$, the longitudinal component of a vector field $A_{\mu}$, 
and the matter field, besides two tensor polarizations and two 
transverse vector components. 
So far the computation of the second-order action of scalar perturbations 
in SVT theories was performed in the flat gauge, 
but the gauge choice from the beginning shows some limitations 
depending on the problems under consideration. 
This motivates us to derive the second-order action of 
scalar perturbations and linear perturbation equations of motion 
without fixing any gauge conditions. 
Our gauge-ready formulation of SVT theories is sufficiently general 
to accommodate Horndeski and GP theories as specific cases. 

The second-order scalar action (\ref{Ss}) consists 
of the Lagrangians ${\cal L}_1^{\rm flat}, {\cal L}_2^{\rm flat}, 
{\cal L}_3^{\rm flat}$ derived for the flat gauge in Ref.~\cite{Kasedark} 
and the new Lagrangians ${\cal L}_{\zeta}, {\cal L}_{E}$ 
arising from the perturbations $\zeta$ and $E$. 
The coefficients of terms in ${\cal L}_{\zeta}, {\cal L}_{E}$ 
can be expressed by using those appearing in 
${\cal L}_1^{\rm flat}, {\cal L}_2^{\rm flat}, {\cal L}_3^{\rm flat}$ 
as well as the coefficients present in the second-order actions 
of tensor and vector perturbations. 
This means that the choice of flat gauge does not lose any physical 
content for the purpose of studying the evolution of scalar perturbations.
As we observe in Eqs.~(\ref{L1})-(\ref{L3}), however, the quantities 
$q_t$ and $c_t^2$ relevant to the stability conditions of tensor perturbations 
do not explicitly appear in ${\cal L}_1^{\rm flat}, {\cal L}_2^{\rm flat}, 
{\cal L}_3^{\rm flat}$, while this is not the case for 
${\cal L}_{\zeta}, {\cal L}_{E}$.
If we choose gauges in which the perturbation $\zeta$ does not vanish, 
this allows one to identify contributions to scalar perturbations arising from 
the tensor sector much easier.

In Sec.~\ref{gaugesec}, we studied the issue of gauge transformations and
constructed a number of gauge-invariant variables associated 
with scalar perturbations. 
In SVT theories, the time-dependent temporal vector component 
$A_0$ contributes to the background evolution 
besides the scalar field $\phi$, so the dynamics is effectively 
described by a multi-scalar system with an adiabatic 
velocity (\ref{dotsigma}).
In Eq.~(\ref{Rpc}), we introduced gauge-invariant curvature perturbations 
${\cal R}_{\phi}$ and ${\cal R}_{\psi}$ associated with the 
scalar perturbation $\delta \phi$ and the longitudinal scalar perturbation $\psi$. 
The total curvature perturbation (\ref{calR}), 
which incorporates both the perturbations $\delta \phi$ and $\psi$, 
can be used for the computation of primordial scalar power 
spectrum generated during inflation. 
We also obtained the general relation between two gauge-invariant 
gravitational potentials $\Psi$ and $\Phi$ in the form (\ref{anire}). 
In Horndeski theories and SVT theories with the couplings (\ref{coup}), 
this relation reduces to the even simpler form (\ref{anire2}).

In Sec.~\ref{stasec}, we derived conditions for avoiding scalar 
ghost and Laplacian instabilities by choosing several different gauges
introduced in Eqs.~(\ref{flat})-(\ref{gauge1}). 
The quantity $q_s$ defined by Eq.~(\ref{qsdef}), whose positivity 
is required for the absence of scalar ghosts, contains the 
common factor $q_t {\cal D}$ irrespective of the gauge choices. 
Provided that the tensor ghost is absent ($q_t>0$) and that the determinant 
${\cal D}$ appearing in the closed-form background equations 
of motion remains positive, the scalar ghost does not appear. 
By computing the scalar propagation speed squares 
$c_{s1}^2$ and $c_{s2}^2$ in several different gauges, 
we explicitly showed that they are gauge-independent quantities.

In Sec.~\ref{appsec}, we applied our general results of Sec.~\ref{scasec} 
to nonsingular bouncing and genesis cosmologies. 
In the flat gauge the quantity $q_s$ is proportional to 
$H^2 q_t {\cal D}$, so it vanishes at the bounce ($H=0$). 
This originates from the inappropriate gauge choice in which 
the curvature perturbations ${\cal R}_{\phi}$ and 
${\cal R}_{\psi}$ vanish at $H=0$. 
If we choose appropriate gauges in which ${\cal R}_{\phi}$ 
and ${\cal R}_{\psi}$ are well defined at the bounce, 
$q_s$ does not cross 0.
We also studied the possibility for realizing nonsingular 
bouncing/genesis cosmologies under the condition that the 
product $q_t c_t^2$ does not asymptotically approach 0 
and showed that, in GP and SVT theories, the existence of intrinsic 
vector modes (with $q_v>0$) can evade the no-go statement for 
the absence of stable nonsingular cosmologies made in Horndeski theories.

In Sec.~\ref{conmodelsec}, we computed observables associated 
with the growth of nonrelativistic matter perturbations for SVT theories 
in which the scalar and vector fields are responsible for the late-time 
cosmic acceleration. By choosing the unitary gauge and using 
the quasi-static approximation on sub-horizon scales, we obtained 
the effective gravitational coupling $\mu$ and the gravitational 
slip parameter $\eta$ in the forms (\ref{muf}) and (\ref{etaf}), respectively.
The quantity $\mu$ can be also expressed as Eq.~(\ref{muex}), where 
$\mu_0$ depends on $q_t, c_t^2, q_v$. 
The second term in Eq.~(\ref{muex}), which depends on 
$q_t, c_t^2, q_v, q_s, c_{s1}^2, c_{s2}^2$, corresponds to 
the interaction of matter with tensor, vector, scalar sectors. 
Unlike the choice of flat gauge \cite{Kasedark}, 
this separation into tensor, vector, scalar contributions is convenient 
to study the cases in which the gravitational 
interaction is stronger or weaker than that in GR. 

In SVT theories satisfying the condition $c_t^2=1$, 
the quantities $\mu_0, \mu_1, \mu_2$ in Eq.~(\ref{muex})
simply reduce to Eq.~(\ref{mu012}).
In cubic functions of the forms (\ref{f3phi}),  
$\mu$ and $\eta$ can be expressed as 
Eqs.~(\ref{muf2}) and (\ref{etaf2}), respectively.
These expressions are analogous to those in Horndeski theories with $c_t^2=1$, 
but the important difference is that the vector mass squared $\alpha_2$ 
appears in SVT theories. 
For $\alpha_2<0$ the gravitational interaction can be 
weaker than that in GR, but in this case it is nontrivial 
to construct consistent dark energy models 
in which the scalar ghost never appears.
It will be of interest to study such a possibility further 
to distinguish SVT theories from Horndeski theories. 

Our gauge-ready formulation of scalar perturbations can be 
directly applicable to the construction of concrete 
bouncing/genesis models in the framework of GP and 
SVT theories. In such cases, the intrinsic vector modes 
should play crucial roles for realizing stable  
solutions. In the context of inflationary cosmology, 
it will be interesting to study 
the effect of the vector field on the primordial power spectrum 
of total curvature perturbations ${\cal R}$.
These issues are left for future works.

\section*{Acknowledgements}

We are grateful to Antonio De Felice and 
Atsushi Naruko for useful discussions. 
LH thanks financial support from Dr.~Max R\"ossler, 
the Walter Haefner Foundation and the ETH Zurich
Foundation.  
RK is supported by the Grant-in-Aid for Young 
Scientists B of the JSPS No.\,17K14297. 
ST is supported by the Grant-in-Aid 
for Scientific Research Fund of the JSPS No.~16K05359 and 
MEXT KAKENHI Grant-in-Aid for 
Scientific Research on Innovative Areas ``Cosmic Acceleration'' (No.\,15H05890).

\appendix

%
\section{Coefficients in the second-order action of scalar perturbations}
\label{coeff}

The coefficients $D_{1,\cdots,10}$ and $w_{1,\cdots,8}$ 
appearing in the background Eqs.~(\ref{back2}), (\ref{back3}),  
(\ref{back5}) and the second-order action of 
scalar perturbations are given by
\ba
&&
D_1=H^3\tp\left(3 G_{5,X_1}+\frac72\tp^2G_{5,X_1X_1}+\frac12\tp^4G_{5,X_1X_1X_1}\right)
+3H^2\left[G_{4,X_1}-G_{5,\phi}+\tp^2\left(4G_{4,X_1X_1}-\frac52G_{5,X_1\phi}\right)\right.
\notag\\
&&\hspace{0.9cm}
\left.+\tp^4\left(G_{4,X_1X_1X_1}-\frac12G_{5,X_1X_1\phi}\right)\right]
-3H\tp\left[G_{3,X_1}+3G_{4,X_1\phi}+\tp^2\left(\frac12G_{3,X_1X_1}+G_{4,X_1X_1\phi}\right)\right]
\notag\\
&&\hspace{0.9cm}
+\frac{1}{2}\left[f_{2,X_1}+2G_{3,\phi}+\tp^2\left(f_{2,X_1X_1}+G_{3,X_1\phi}\right)
+\tp A_0 f_{2,X_1X_2}+\frac{A_0^2}{4}f_{2,X_2X_2}\right]
\,,
\notag\\
&&
D_2=-\left[2(G_{4,X_1}-G_{5,\phi})+\tp^2(2G_{4,X_1X_1}-G_{5,X_1\phi})
+H\tp(2G_{5,X_1}+\tp^2G_{5,X_1X_1})\right]\dot{H}
\notag\\
&&\hspace{0.9cm}
+\left[G_{3,X_1}+3G_{4,X_1\phi}+\tp^2\left(\frac{G_{3,X_1X_1}}{2}+G_{4,X_1X_1\phi}\right)
-2H\tp(3G_{4,X_1X_1}-2G_{5,X_1\phi})\right.
\notag\\
&&\hspace{0.9cm}
\left.-H\tp^3(2G_{4,X_1X_1X_1}-G_{5,X_1X_1\phi})
-H^2\left(G_{5,X_1}+\frac52\tp^2G_{5,X_1X_1}+\frac12\tp^4G_{5,X_1X_1X_1}
\right)
\right]\ddot{\phi}
\notag\\
&&\hspace{0.9cm}
-H^3\tp\left(2G_{5,X_1}+\tp^2G_{5,X_1X_1}\right)
-H^2\left[3(G_{4,X_1}-G_{5,\phi})+5\tp^2\left(G_{4,X_1X_1}-\frac12G_{5,X_1\phi}\right)
+\frac12\tp^4G_{5,X_1X_1\phi}\right]
\notag\\
&&\hspace{0.9cm}
+2H\tp(G_{3,X_1}+3G_{4,X_1\phi})-H\tp^3(2G_{4,X_1X_1\phi}-G_{5,X_1\phi\phi})
+\tp^2\left(\frac12G_{3,X_1\phi}+G_{4,X_1\phi\phi}\right)-G_{3,\phi}-\frac12 f_{2,X_1}\,,
\notag\\
&&
D_3=3\left[G_{4,\phi\phi}+f_{4,\phi\phi}+\tp^2\left(\frac12G_{3,X_1\phi}+G_{4,X_1\phi\phi}\right)
+HA_0f_{5,\phi\phi}-2H\tp(G_{4,X_1\phi}-G_{5,\phi\phi})\right.
\notag\\
&&\hspace{0.9cm}
\left.-H\tp^3(2G_{4,X_1X_1\phi}-G_{5,X_1\phi\phi})
-\frac{H^2\tp^2}{2}\left(3G_{5,X_1\phi}+\tp^2G_{5,X_1X_1\phi}\right)
\right]\dot{H}
\notag\\
&&\hspace{0.9cm}
-\left[\frac12f_{2,X_1\phi}+G_{3,\phi\phi}
+\frac12\tp A_0f_{2,X_1X_2\phi}+\frac{A_0^2}{8}f_{2,X_2X_2\phi}
+\frac12\tp^2(f_{2,X_1X_1\phi}+G_{3,X_1\phi\phi})
-3H\tp(G_{3,X_1\phi}+3G_{4,X_1\phi\phi})\right.
\notag\\
&&\hspace{0.9cm}
\left.
-3H\tp^3\left(\frac12G_{3,X_1X_1\phi}+G_{4,X_1X_1\phi\phi}\right)
+3H^2(G_{4,X_1\phi}-G_{5,\phi\phi})
+3H^2\tp^2(4G_{4,X_1X_1\phi}-\frac52G_{5,X_1\phi\phi})\right.
\notag\\
&&\hspace{0.9cm}
\left.
+3H^2\tp^4\left(G_{4,X_1X_1X_1\phi}-\frac12G_{5,X_1X_1\phi\phi}\right)
+H^3\tp\left(3G_{5,X_1\phi}+\frac72\tp^2G_{5,X_1X_1\phi}
+\frac12\tp^4G_{5,X_1X_1X_1\phi}\right)
\right]\ddot{\phi}
\notag\\
&&\hspace{0.9cm}
-\frac32H^4\tp^2(3G_{5,X_1\phi}+\tp^2G_{5,X_1X_1\phi})
+H^3\left[\frac12A_0(9f_{5,\phi\phi}+A_0^2f_{5,X_3\phi\phi})
-9\tp(G_{4,X_1\phi}-G_{5,\phi\phi})\right.
\notag\\
&&\hspace{0.9cm}
\left.-\tp^3\left(9G_{4,X_1X_1\phi}-\frac72G_{5,X_1\phi\phi}\right)
-\frac12\tp^5G_{5,X_1X_1\phi\phi}\right]
+3H^2\left[2f_{4,\phi\phi}+2G_{4,\phi\phi}+A_0^2f_{4,X_3\phi\phi}\right.
\notag\\
&&\hspace{0.9cm}
\left.+\frac{\dot{A_0}(f_{5,\phi\phi}+A_0^2f_{5,X_3\phi\phi})}{2}
+\tp^2\left(\frac32G_{3,X_1\phi}+3G_{4,X_1\phi\phi}+\frac12G_{5,\phi\phi\phi}\right)
-\tp^4\left(G_{4,X_1X_1\phi\phi}-\frac12G_{5,X_1\phi\phi\phi}\right)\right]
\notag\\
&&\hspace{0.9cm}
-3H\left[\frac{A_0(f_{2,X_2\phi}+4f_{3,\phi\phi})}{4}-A_0\dot{A_0}f_{4,X_3\phi\phi}
+\tp\left(\frac{1}{2}f_{2,X_1\phi}+G_{3,\phi\phi}\right)
-\tp^3\left(\frac12G_{3,X_1\phi\phi}+G_{4,X_1\phi\phi\phi}\right)\right]
\notag\\
&&\hspace{0.9cm}
-\frac14\tp^2(2f_{2,X_1\phi\phi}+\dot{A_0}f_{2,X_1X_2\phi}+2G_{3,\phi\phi\phi})
-\frac{\tp A_0}{4}\left[f_{2,X_2\phi\phi}
+\frac12\dot{A_0}(4f_{2,X_1X_3\phi}+f_{2,X_2X_2\phi})\right]
\notag\\
&&\hspace{0.9cm}
-\dot{A_0}\left(f_{3,\phi\phi}-A_0^2\tilde{f}_{3,\phi\phi}
+\frac{f_{2,X_2\phi}+A_0^2f_{2,X_2X_3\phi}}{4}\right)+\frac{f_{2,\phi\phi}}{2}\,,
\notag\\
&&
D_4=-H^3\tp^2\left(15G_{5,X_1}+10\tp^2G_{5,X_1X_1}+\tp^4G_{5,X_1X_1X_1}\right)
+3H^2\left[A_0(f_{5,\phi}-A_0^2f_{5,X_3\phi})-6\tp(G_{4,X_1}-G_{5,\phi})\right.
\notag\\
&&\hspace{0.9cm}
\left.-\tp^3(12G_{4,X_1X_1}-7G_{5,X_1\phi})-\tp^5(2G_{4,X_1X_1X_1}-G_{5,X_1X_1\phi})\right]
+3H\left[2(f_{4,\phi}+G_{4,\phi})-2A_0^2f_{4,X_3\phi}\right.
\notag\\
&&\hspace{0.9cm}
\left.+\tp^2(3G_{3,X_1}+8G_{4,X_1\phi})+\tp^4(G_{3,X_1X_1}+2G_{4,X_1X_1\phi})\right]
-\tp^3(f_{2,X_1X_1}+G_{3,X_1\phi})-\frac12\tp^2A_0f_{2,X_1X_2}
\notag\\
&&\hspace{0.9cm}
-\tp (f_{2,X_1}-A_0^2f_{2,X_1X_3}+2G_{3,\phi})
+\frac12A_0(f_{2,X_2}+A_0^2f_{2,X_2X_3}+4f_{3,\phi}-4A_0^2\tilde{f}_{3,\phi})\,,
\notag\\
&&
D_5=H^3\left[A_0^3 (f_{5,X_3\phi}+A_0^2f_{5,X_3X_3\phi})
-\tp^3(5G_{5,X_1\phi}+\tp^2G_{5,X_1X_1\phi})\right]
+3H^2\left[2(f_{4,\phi}+G_{4,\phi}+A_0^4f_{4,X_3X_3\phi})\right.
\notag\\
&&\hspace{0.9cm}
\left.+\tp A_0 ( f_{5,\phi\phi}-A_0^2f_{5,X_3\phi\phi} )
-\tp^2(4G_{4,X_1\phi}-3G_{5,\phi\phi})-\tp^4(2G_{4,X_1X_1\phi}-G_{5,X_1\phi\phi})\right]
\notag\\
&&\hspace{0.9cm}
-3H\left[2A_0^3(f_{3,X_3\phi}+\tilde{f}_{3,\phi})
-2\tp(f_{4,\phi\phi}-A_0^2f_{4,X_3\phi\phi}+G_{4,\phi\phi})
-\tp^3(G_{3,X_1\phi}+2G_{4,X_1\phi\phi})\right]
\notag\\
&&\hspace{0.9cm}
-\tp^2(f_{2,X_1\phi}+G_{3,\phi\phi})+2\tp A_0(f_{3,\phi\phi}-A_0^2\tilde{f}_{3,\phi\phi})
+f_{2,\phi}+A_0^2f_{2,X_3\phi}
\,,
\notag\\
&&
D_6=H^2\tp^2(3G_{5,X_1}+\tp^2G_{5,X_1X_1})
-2H\left[A_0f_{5,\phi}-2\tp(G_{4,X_1}-G_{5,\phi})-\tp^3(2G_{4,X_1X_1}-G_{5,X_1\phi})\right]
\notag\\
&&\hspace{0.9cm}
-\tp^2(G_{3,X_1}+2G_{4,X_1\phi})-2(f_{4,\phi}+G_{4,\phi})
\,,
\notag\\
&&
D_7=H^3\tp^2(3G_{5,X_1}+\tp^2G_{5,X_1X_1})
-H^2\left[A_0(3f_{5,\phi}+A_0^2f_{5,X_3\phi})
-6\tp(G_{4,X_1}-G_{5,\phi})-2\tp^3(3G_{4,X_1X_1}-2G_{5,X_1\phi})\right]
\notag\\
&&\hspace{0.9cm}
-H\left[2(f_{4,\phi}+2A_0^2f_{4,X_3\phi}+G_{4,\phi})-2A_0\tp f_{5,\phi\phi}
+\tp^2(3G_{3,X_1}+10G_{4,X_1\phi}-2G_{5,\phi\phi})\right]
\notag\\
&&\hspace{0.9cm}
+\tp(f_{2,X_1}+2f_{4,\phi\phi}+2G_{3,\phi}+2G_{4,\phi\phi})
+\frac12A_0(f_{2,X_2}+4f_{3,\phi})
\,,
\notag\\
&&
D_8=-\frac{2\tp D_1+D_4+3HD_6}{A_0}\,,
\notag\\
&&
D_9=-H^3A_0^2(3f_{5,X_3\phi}+A_0^2f_{5,X_3X_3\phi})
-3H^2\left[2A_0(f_{4,X_3\phi}+A_0^2f_{4,X_3X_3\phi})
-\tp(f_{5,\phi\phi}+A_0^2f_{5,X_3\phi\phi})\right]
\notag\\
&&\hspace{0.9cm}
+6HA_0\left[A_0(f_{3,X_3\phi}+\tilde{f}_{3,\phi})+\tp f_{4,X_3\phi\phi}\right]
-\tp\left(\frac12f_{2,X_2\phi}+2f_{3,\phi\phi}-2A_0^2\tilde{f}_{3,\phi\phi}\right)
-A_0f_{2,X_3\phi}
\,,
\notag\\
&&
D_{10}=-2\dot{H}f_{5,\phi}-H^2 \left(3f_{5,\phi}+A_0^2f_{5,X_3\phi} \right)
-2HA_0 \left(2f_{4,X_3\phi}+\dot{A_0}f_{5,X_3\phi} \right)
-2\dot{A_0}f_{4,X_3\phi}+2f_{3,\phi}+\frac{1}{2} f_{2,X_2}\,,
\ea
and 
\ba
&&
w_1=-H^2\left[A_0^3(f_{5,X_3}+A_0^2f_{5,X_3X_3})-\tp^3(5G_{5,X_1}+\tp^2G_{5,X_1X_1})\right]
-2H\left[2(f_4+A_0^4f_{4,X_3X_3}+G_4)\right.
\notag\\
&&\hspace{0.9cm}
\left.+A_0\tp(f_{5,\phi}-A_0^2f_{5,X_3\phi})-\tp^2(4G_{4,X_1}-3G_{5,\phi})
-\tp^4(2G_{4,X_1X_1}-G_{5,X_1\phi})\right]
\notag\\
&&\hspace{0.9cm}
-\tp^3(G_{3,X_1}+2G_{4,X_1\phi})-2\tp(f_{4,\phi}-A_0^2f_{4,X_3\phi}
+G_{4,\phi})+2A_0^3(f_{3,X_3}+\tilde{f}_3)
\,,
\notag\\
&&
w_2=w_1+2Hq_t-\tp D_6\,,
\notag\\
&&\hspace{0.51cm}
=A_0 \left[
-H^2A_0^2(3f_{5,X_3}+A_0^2f_{5,X_3X_3})
-2H\left[2A_0(f_{4,X_3}+A_0^2f_{4,X_3X_3})
-\tp(f_{5,\phi}+A_0^2f_{5,X_3\phi})\right]
+2A_0\tp f_{4,X_3\phi}
\right.\notag\\
&&\hspace{0.8cm}
\left.
+2A_0^2( f_{3,X_3}+\tilde{f}_3) \right]\,,
\notag\\
&&
w_3=-2A_0^2q_v\,,
\notag\\
&&
w_4=w_5-H^3\left[3A_0^3(2f_{5,X_3}+A_0^2f_{5,X_3X_3})
-\tp^3\left(15G_{5,X_1}+\frac{13}{2}\tp^2G_{5,X_1X_1}+\frac12\tp^4G_{5,X_1X_1X_1}\right)\right]
-3H^2\left[2(f_4+G_4)\right.
\notag\\
&&\hspace{0.9cm}
\left.+A_0^2\left(2f_{4,X_3}+4A_0^2f_{4,X_3X_3}-3A_0\tp f_{5,X_3\phi}\right)
-\tp^2(7G_{4,X_1}-6G_{5,\phi})-\tp^4\left(8G_{4,X_1X_1}-\frac92G_{5,X_1\phi}\right)\right.
\notag\\
&&\hspace{0.9cm}
\left.-\tp^6\left(G_{4,X_1X_1X_1}-\frac12G_{5,X_1X_1\phi}\right)\right]
+3H\left[2A_0^3(f_{3,X_3}+\tilde{f}_3)-2\tp(f_{4,\phi}-2A_0^2f_{4,X_3\phi}+G_{4,\phi})\right.
\notag\\
&&\hspace{0.9cm}
\left.-\tp^3(2G_{3,X_1}+5G_{4,X_1\phi})-\tp^5\left(\frac12G_{3,X_1X_1}+G_{4,X_1X_1\phi}\right)\right]
+\frac12\tp^4(f_{2,X_1X_1}+G_{3,X_1\phi})
\notag\\
&&\hspace{0.9cm}
+\tp^2\left(\frac12f_{2,X_1}-A_0^2f_{2,X_1X_3}-\frac18A_0^2f_{2,X_2X_2}+G_{3,\phi}\right)
-\frac12A_0\tp\left(f_{2,X_2}+A_0^2f_{2,X_2X_3}+4f_{3,\phi}-4A_0^2\tilde{f}_{3,\phi}\right)
\,,
\notag\\
&&
w_5=\frac12H^3A_0^3(3f_{5,X_3}+6A_0^2f_{5,X_3X_3}+A_0^4f_{5,X_3X_3X_3})
+3H^2A_0\left[A_0^3(3f_{4,X_3X_3}+A_0^2f_{4,X_3X_3X_3})\right.
\notag\\
&&\hspace{0.9cm}
\left.+\frac12\tp(f_{5,\phi}-2A_0^2f_{5,X_3\phi}-A_0^4f_{5,X_3X_3\phi})\right]
-3HA_0^3\left[f_{3,X_3}+\tilde{f}_3+A_0^2(f_{3,X_3X_3}+\tilde{f}_{3,X_3})
+A_0\tp f_{4,X_3X_3\phi}\right]
\notag\\
&&\hspace{0.9cm}
+\frac18A_0^2\tp^2f_{2,X_2X_2}-\frac14A_0\tp\left[f_{2,X_2}+4f_{3,\phi}
-2A_0^2(f_{2,X_2X_3}-2\tilde{f}_{3,\phi}+2f_{3,X_3\phi})+4A_0^4\tilde{f}_{3,X_3\phi}\right]
+\frac12A_0^4f_{2,X_3X_3}
\,,
\notag\\
&&
w_6=-\frac{w_1-\tp D_6+2Hq_t}{A_0}-4H\left(2A_0f_{4,X_3}-\tp f_{5,\phi}
+HA_0^2f_{5,X_3}\right)\,,
\notag\\
&&\hspace{0.51cm}
=-H^2A_0^2(f_{5,X_3}-A_0^2f_{5,X_3X_3})
-2H\left[2A_0(f_{4,X_3}-A_0^2f_{4,X_3X_3})
-\tp (f_{5,\phi}-A_0^2f_{5,X_3\phi})\right]
-2 A_0\tp f_{4,X_3\phi}\notag\\
&&\hspace{0.9cm}
-2A_0^2( f_{3,X_3}+\tilde{f}_3)\,,
\notag\\
&&
w_7=-2\dot{H} \left(2f_{4,X_3}+HA_0f_{5,X_3} \right)
-H^2\left[\frac{\tp(3f_{5,\phi}+A_0^2f_{5,X_3\phi})}{A_0}
+\dot{A_0} \left(f_{5,X_3}+A_0^2f_{5,X_3X_3} \right)\right]\notag\\
&&\hspace{0.9cm}
-4H \left( \tp f_{4,X_3\phi}+A_0\dot{A_0}f_{4,X_3X_3} \right)
+2\dot{A_0} \left( f_{3,X_3}+\tilde{f}_3 \right)
+\frac{\tp( f_{2,X_2}+4f_{3,\phi})}{2A_0}\,,
\notag\\
&&
w_8=3Hw_1-2w_4-\tp D_4\,.
\ea
We note that Eqs.~(\ref{back1}) and (\ref{back4}) were used 
for the derivation of these coefficients. 

\section{$\alpha_{\rm T}$ and $\alpha_{\rm K}$}
\label{aKaT}

Besides the quantities $\alpha_{\rm M}$ and $\alpha_{\rm B}$ given in 
Eqs.~(\ref{alM}) and (\ref{EFTdef}), we define
the following dimensionless quantities: 
\ba
&&\alpha_{\rm T}\equiv\frac{1}{q_t}\left[2A_0^2f_{4,X_3}+2\tp^2G_{4,X_1}
-2\tp A_0f_{5,\phi}-2\tp^2G_{5,\phi}+A_0^2(HA_0-\dot{A_0})f_{5,X_3}
+\tp^2(H\tp-\ddot{\phi})G_{5,X_1}\right]\,,\label{aT} \\
&&\alpha_{\rm K}\equiv6+12\alpha_{\rm B}
+\frac{2(w_4+4w_5+2w_8)}{H^2q_t}\,.
\label{aK}
\ea
After switching off the vector field, Eqs.~(\ref{aT}) and (\ref{aK}) reduce to those 
in Horndeski theories  introduced in Ref.~\cite{Bellini}.  
The quantity $\alpha_{\rm T}$ represents the deviation of $c_t^2$ from 
that of light, i.e.,  $c_t^2=1+\alpha_{\rm T}$, while $\alpha_{\rm K}$ corresponds 
to the kinetic term for scalar perturbations.
The matrix component $K_{22}$ given in Eq.~(\ref{Kcom}), which is 
computed in the unitary gauge, can be 
simply expressed in terms of $q_t$, $\alpha_{\rm B}$ 
and $\alpha_{\rm K}$, as 
\be
K_{22}=\frac{q_t(\alpha_{\rm K}+6\alpha_{\rm B}^2)}{2(1+\alpha_{\rm B})^2}\,.
\ee
%


\end{document}